\documentclass[fleqn,10pt]{wlscirep}
\usepackage[utf8]{inputenc}
\usepackage[T1]{fontenc}

\title{Theory of Quantum Path Computing with Fourier Optics and Future Applications for Quantum Supremacy,  Neural Networks and Nonlinear Schr{\"o}dinger Equations}

\author[1,*]{Burhan Gulbahar}
\affil[1]{Department of Electrical and Electronics Engineering, Ozyegin University, 34794 Istanbul, Turkey}
\affil[*]{burhan.gulbahar@ozyegin.edu.tr \\

The fully edited version of the published article is available in  \href{https://www.nature.com/articles/s41598-020-67364-0}{https://www.nature.com/articles/s41598-020-67364-0} with DOI \href{https://doi.org/10.1038/s41598-020-67364-0}{https://doi.org/10.1038/s41598-020-67364-0}. This is the author accepted copy of the original article.}

\usepackage{graphicx}% Include figure files
\usepackage{dcolumn}% Align table columns on decimal point
\usepackage{bm}% bold math
\usepackage{url}
\usepackage{epstopdf}
\usepackage{amsmath} 
\usepackage{relsize}
\usepackage{multirow}
\usepackage{physics}
\usepackage{makecell}
\usepackage{hyperref}
\hypersetup{
    colorlinks=true,
    linkcolor=blue,
    filecolor=magenta,      
    urlcolor=cyan,
    citecolor=blue,
}

\usepackage{amsmath}
\usepackage[cmintegrals]{newtxmath}
\usepackage{mciteplus}
\usepackage{cite}
\usepackage{graphicx}% Include figure files
\usepackage{dcolumn}% Align table columns on decimal point
\usepackage{bm}% bold math
\usepackage{relsize}
\usepackage{multirow}
\usepackage{makecell}
\usepackage{color}
\usepackage{mathtools}
\makeatletter
\@fleqnfalse
\@mathmargin\@centering
\makeatother
\usepackage{mathdots}% http://ctan.org/pkg/mathdots

\usepackage{braket}
\usepackage{makecell}
\usepackage{relsize}

\begin{abstract}
The scalability, error correction and practical problem solving {\color{black}are important} challenges for quantum computing (QC) as more emphasized by quantum supremacy (QS) experiments. Quantum path computing (QPC), recently introduced for linear optic based QCs (LOQCs) as an unconventional design, {\color{black}targets to obtain} scalability and practical problem solving. {\color{black}It samples} the intensity from {\color{black}the} interference of exponentially increasing number of propagation paths {\color{black} obtained in multi-plane diffraction (MPD) of classical particle sources}. {\color{black}  QPC exploits MPD based quantum temporal correlations of the paths and freely entangled projections at different time instants, for the first time, with the classical light source and intensity measurement while not requiring photon interactions or single photon sources and receivers.} In this article, photonic QPC is defined, theoretically modeled and numerically analyzed for arbitrary Fourier optical or quadratic phase set-ups while utilizing both Gaussian and Hermite-Gaussian source laser modes. Problem solving capabilities  {\color{black}already} including {\color{black}partial sum} of Riemann theta functions are {\color{black}extended}. Important future applications, {\color{black} implementation challenges}  and open issues {\color{black} such as universal computation and quantum circuit implementations determining the scope of QC capabilities} are {\color{black}discussed.} The {\color{black}applications} include QS experiments {\color{black}reaching} more than $2^{100}$ Feynman paths, quantum neuron implementations and solutions of nonlinear Schr{\"o}dinger equation. 
\end{abstract}

\begin{document}

\flushbottom
\maketitle
 
\thispagestyle{empty}

The scalability of quantum resources including qubits and quantum gates, improved error correction capabilities and practical problem solving ability are the most important challenges for modern quantum computing (QC). Recent quantum supremacy experiments (QS) of Google as a success milestone for the human history of computing emphasize the importance of these properties in their set-up with $53$ qubits and $20$ cycles reaching Hilbert space size of \textbf{$\approx 2^{66}$} Feynman paths making it significantly difficult to classically calculate their result \cite{arute2019quantum, cho2019google}. QS experiments target to show the computational capability of QCs such that feasible computations obtained with QCs require significant resources to perform with classical computers \cite{preskill2018quantum}. In the global and highly competitive race including technology giants, a wide variety of but quite complex hardware architectures are used. For example, Google, IBM and Rigetti Computing use superconducting circuits while Microsoft using topological anions generated by frozen nanowires, Ion-Q using ion traps at room temperature and  D-Wave using quantum annealing technology \cite{gomes2018quantum}.  On the other hand, both the challenges of scalability and practical problem solving capability continue to exist such that QCs that can show significant advantages in practical problem solving compared to conventional computers require much more resources such as  thousands of logical qubits and hundreds of thousands physical qubits \cite{childs2018toward}. Therefore, building QC system architectures which are more tolerant to noise and decoherence combined with capabilities  of error correction, practical problem solving, low hardware complexity and resource scalability is significantly important for near-term advantages of QC. 

\begin{figure}[!t]
\centering
\includegraphics[width=3.5in]{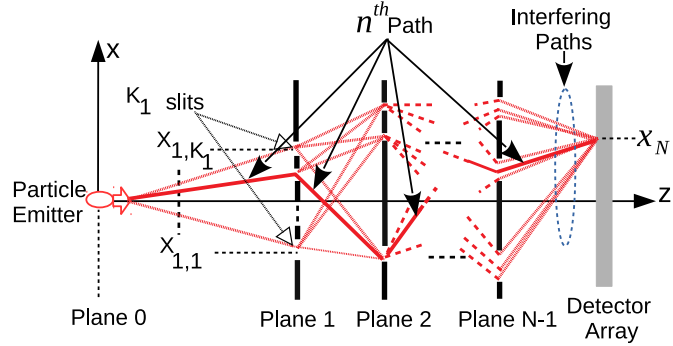}  
\caption{QPC set-up composed of $N-1$ diffraction planes, FSP between the planes and a single sensor plane on which exponentially increasing number of propagation paths interfere \cite{gulbahar2019quantumpath}.}
\label{Fig1}
\end{figure} 

Linear optic (LO) based QCs (LOQCs) have a special place among the existing QC architectures due to easy manipulation of photons, unique features of photons not interacting with the environment in terms of decoherence, working at room temperature and maturity in classical optics for centuries \cite{kok2007linear, li2015resource, carolan2015universal}. For example, boson sampling \cite{aaronson2011computational}  as a candidate for LOQC based QS promises experimental implementation of QS in the future while the recent experimental achievements in Ref. \citen{wang2019boson} improve the complexity of the solved problems gradually reaching Hilbert space size of \textbf{$\approx 2^{48}$}. However, the existing LOQC candidates for QS have still the fundamental challenges of scalability of the resources, e.g., the requirement of single photon sources and receivers, and practical problem solving capability, e.g., matrix permanents in boson sampling. Furthermore, multi-photon entanglement resources and quantum circuits are challenging to create due to the difficulty in the interaction of photons with each other.  Multi-plane diffraction (MPD) based QC system denoted by quantum path computing (QPC) as shown in Fig. \ref{Fig1} {\color{black}is} recently proposed in Ref.  \citen{gulbahar2019quantumpath} as  one of the simple LOQC architectures {\color{black}targeting to realize} scalability and practical problem solving capabilities. Sampling from the interference of exponentially increasing number of propagation paths {\color{black}and the freely entangled projections at different time instants \cite{gulbahar2018quantum} are} utilized for QC purposes, {\color{black} for the first time, by exploiting coherent and classical sources such as standard laser sources}.  Exponentially increasing number of Feynman paths with respect to the given amount of slits and diffraction planes makes the classical calculation of the interference output significantly hard \cite{gulbahar2019quantumpath}. The unique form of {\color{black}temporal correlation freely available} among the exponentially increasing number of Feynman paths {\color{black}in the MPD set-up} is denoted as  quantum path entanglement (QPE)  in Ref. \citen{gulbahar2018quantum} as a novel resource to exploit for QC based on the coherence and superposition of the classical light source \cite{sup2, sup3}.  {\color{black} In fact, a quantum mechanical propagator for photons with the form of the classical Fresnel diffraction integral is verified in Ref. \citen{santos2018huygens}  such that  the classical intensity of the field is  proportional to the probability density of photon detection for the position observable transversal to the propagation and in the limit for large number of quanta. Therefore, FO based set-up exploits classical light source and its intensity measurement while exploiting quantum temporal correlations in a unique MPD design.} QPC promises a significant alternative to cope with the fundamental challenges of scalability of multi-photon entanglement resources and the complex requirements of  single photon sources and detection mechanisms observed in conventional linear optical QC systems \cite{carolan2015universal, wang2016experimental, wang2018multidimensional, wang2018toward}.  It has  important all-in-one advantages combining the utilization of the classical sources, i.e., either fermion or boson, hardware simplicity based on diffraction slits and detection with conventional photon counting  intensity detectors without requiring simultaneous detection of multiple photons in multiple registers \cite{aaronson2011computational}. 
   
Besides that, another QC architecture related to multi-slit structures is denoted as duality computer (DC) which exploits duality parallelism for performing different gate operations on the sub-wave functions through sub-waves corresponding to each slit  \cite{dual1, dual2, dual3, dual4} while utilized in various machine learning \cite{dual5} and photonic chip applications \cite{dual6}. The quantum wave divider (QWD) divides wave functions into sub-waves. In addition, sub-sub-waves can also be obtained in a multi-level QWD while quantum wave combiner (QWC) is utilized to combine the waves after performing operations on each path. It is firstly required as an open issue to  explicitly model the theoretical computational complexity and QC capabilities of a multi-level version of DC with respect to the targeted QWD/QWC configurations before comparing with MPD design.  For example, combined operations of uniform QWD/QWC in a DC leaving the state unchanged is not comparable with MPD modifying the state with diffractions as time evolves.  Single level complexity  discussions in the DC related literature include the requirement of an extra qudit \cite{dual2, dual3}  for simulating DC device with an ordinary quantum computer which is not comparable with the tensor product structure of MPD requiring multiple qudits.  The division of each wave into a sub-wave and then into a sub-sub-wave could require an exponentially increasing number of slit resources making it challenging as discussed next for Ref. \citen{dragoman}.  On the other hand, DC based architectures have a long history and maturity along two decades with many capabilities and applications including practical database search without requiring extra resources \cite{dual1} verified with quantum circuit modeling in Ref. \citen{dual2}. DC utilizes linear combination of unitaries (LCU) for QC while becoming one of the five major techniques for designing quantum algorithms \cite{r1},  in multi-party secure computation \cite{r2} and HHL quantum algorithm \cite{r3}.  

Furthermore, simple optical setups exploiting wave-particle duality and interferometers have the cost of exponential complexity of resources either in time, space or energy domains to achieve QC advantages as discussed in Ref. \citen{gulbahar2019quantumpath}. {\color{black}For example,} wave particle (WP) computer \cite{wpduality} {\color{black}exploits} full optical interconnections of an $N \times N$ input signal array with an $N \times N$ output signal creating $N^4$ channels with a tensor product. WP computer also  utilizes a filter array between the input and output to increase the number of connections in an additive manner with respect to the connections in each inter-planar region. Such  architectures, including also Ref. \citen{dragoman} for a slit based modeling, provide  advantages of parallelism compared with classical models without exploiting temporal correlations of quantum histories and their tensor product structure \cite{gulbahar2018quantum}.  They utilize the tensor product only for a single inter-planar propagation, i.e.,  a single measurement plane directly detecting propagation from the input array. The rich set of two and three dimensional alternative optical interconnection architectures and opto-electronic computing are discussed in detail  in Ref. \citen{haldunoptical1} by also including multi-stage interconnection topology. Analog Fourier optics {\color{black}(FO)} and  its digital equivalent, i.e., digital {\color{black}FO} architectures composed of smart pixel arrays of two-dimensional electronic processing units connected with optical interconnections,  exploit speed and parallelism advantages of the optical design \cite{haldunoptical2}. Furthermore, programmable directed logic networks are discussed in Ref. \citen{wplogic} by emphasizing the energy efficiency of optical architectures.

On the other hand, QPC formulation is performed for electron based set-up in Ref. \citen{gulbahar2019quantumpath} while theoretical studies {\color{black}modeling}  QPE   in Ref. \citen{gulbahar2018quantum}  and classical optical communications in Ref. \citen{ gulbahar2019quantumspatial} formulate free space propagation (FSP) of light. {\color{black}They do not   generalize} to  arbitrary set-ups of {\color{black}FO}, i.e., first order centered optical or quadratic-phase systems including arbitrary sections of free space, thin lenses, graded index media and spatial filters \cite{ozaktas2001fractional} and mathematically characterized as linear canonical transforms (LCTs)\cite{healy2015linear}. LCTs are linear integral transforms including the Fresnel and fractional Fourier transform (FRFT),  scaling, chirp multiplication and some other operations as special cases while being equivalent to  spatial distribution of light in phase-space optics for quadratic-phase systems \cite{ozaktas2001fractional}. Besides that, previous MPD studies utilize Gaussian sources without extending to Hermite-Gaussian (HG) beams compatible with the standard laser sources within the paraxial approximation \cite{pampaloni2004gaussian}. Photonic QPC formulation is not available  while important applications of QPC other than the partial sum of Riemann theta function (RTF) and period finding presented in Ref. \citen{gulbahar2019quantumpath} are not discussed and theoretically analyzed yet. 

Diffractive and phase space optics are also getting attention in quantum technologies with periodic single plane diffraction  for implementing  quantum logic gates using quantum Talbot effect \cite{sawada2018experimental}, for testing D-dimensional (qudit) Bell inequality with free space entangled quantum carpets \cite{barros2017free} and for the evaluation of entanglement over the entire transverse field distribution of the photons \cite{tasca2018testing} while without any discussion regarding the MPD based advantages. Proposed theoretical modeling and system design of photonic QPC with widely available optical components, e.g., thin lenses, free space and diffraction planes as a form of spatial filtering, provide a unique opportunity to exploit conventional FO for QC.  The large amount of theoretical and experimental maturity in FO since the last century is combined with MPD based system design to realize  scalable and low complexity QC systems with important capabilities and global resources for efficient implementation and development.

In this article, QPC set-up is defined, theoretically modeled and numerically analyzed for {\color{black}FO} with arbitrary LCTs between diffraction planes. QPC system exploiting diffraction in an unconventional manner maintains {\color{black}photonic advantages including decoherence and noise} while avoids the need to interact with multiple photons by eliminating  many problems encountered in multi-photon entanglement and circuit implementations. Furthermore, the quantum nature of FO is discussed based on the  experimental   \cite{rengaraj2018, santos2018huygens} and theoretical \cite{sawant2014nonclassical} studies   verifying the validity of Fresnel diffraction formulation for quantum optical propagation.  Classical monochromatic light sources of both Gaussian and Hermite-Gaussian (HG) beams are utilized compatible with the standard laser sources within the paraxial approximation \cite{pampaloni2004gaussian}.  LCT based design which provides more flexibility is numerically compared with FSP in terms of improvement on the detection efficiency and the interference complexity defined with the magnitudes of the interfering paths and negative volume of Wigner distribution function \cite{gulbahar2019quantumpath, kenfack2004negativity}.  

{\color{black}Important} future applications of photonic QPC are, for the first time, introduced and theoretically modeled in an introductory and brief manner. These include the feasibility of QS experiments compared with alternative technologies, adapting certified random number generation protocols for the photonic QPC architecture \cite{aaronson2019, whyte2019,  bouland2019complexity, brakerski2018cryptographic}, quantum neural network (QNN) implementations and making the solutions of nonlinear Schr{\"o}dinger equation (NLSE) easier. The detailed modeling and utilization of photonic QPC for these applications are presented as open issues.
 
The potential of QS experiments with photonic QPC is presented in this article to reach {\color{black}more than  $2^{100}$ Feynman paths in a scalable set-up}  with several tens of diffraction planes while requiring experimental implementations {\color{black}for better modeling and verifying} the scalability for large scale QPC set-ups.   {\color{black}A feasible method is proposed}  to exponentially increase the number of Feynman paths with the cost of  linearly increasing  number of planes  {\color{black}and slits allowing to obtain} significantly large Hilbert space.    However, it is an open issue to verify  QS capability both complexity theoretically and experimentally based on the promising results in Ref. \citen{gulbahar2019quantumpath} and the modeling in this article {\color{black} such as by performing analogous modeling and experiments} in Refs. \citen{aaronson2011computational, wang2019boson} and  \citen{aaronson2016} achieved for Boson sampling.   Moreover, QPC with Gaussian sources results in unique mathematical forms of wave functions on the sensor plane in  (\ref{Eq_new_7}) to be exploited for the solutions of the partial sum of RTF \cite{riemann1857theorie, deconinck2004computing, mumford2007tata, osborne2002nonlinear},  period finding \cite{nielsen2002quantum} and Diophantine approximation \cite{lagarias1985computational} similar to the algorithms and methods in Ref. \citen{gulbahar2019quantumpath} but with much more design flexibility due to LCTs, diversity of the tools and maturity in the science of FO.  HG sources result in different forms in (\ref{Eq_new_11}) and (\ref{Eq_new_13}) while closely related to the standard RTF  form and requiring future studies to exploit for the solutions of numerical problems in various scientific disciplines.  {\color{black} On the other hand, open issues and challenges for FO based QPC design are discussed to determine the scope of the proposed design for QC purposes, e.g., universal quantum computation capability, implementations of quantum circuit gates and basic search algorithms such as Grover search.} 

Neural networks (NNs) exploiting the quantum advantages, i.e., QNNs, improve the capabilities of classical NNs with quantum interference and superposition for deep learning applications \cite{lecun2015deep} in various disciplines \cite{cao2017quantum}.  On the other hand, linear and unitary framework of quantum mechanics results in the challenges of implementing non-linear and dissipative dynamics of classical neural networks \cite{cao2017quantum}. The state-of-the-art neuron implementations utilize various methods to introduce non-linearity including quantum measurements \cite{cao2017quantum, tacchino2019artificial}. The quantum interference among the exponentially increasing number of paths and the entanglement denoted as  QPE in Ref. \citen{gulbahar2018quantum} are promising  for designing and practically implementing novel design of QNNs. QPC set-up has inherently nonlinear formulation with respect to slit positions to encode the input and {\color{black}it} operates on the quantum superposition of the inputs. Besides that, implementations of diffractive NNs utilizing single-layer \cite{brunner2015reconfigurable, bueno2018reinforcement} and all-optical multi-layer diffractive architectures \cite{lin2018all} do not exploit interference among the paths or quantum domain advantages. {\color{black}Photonic} QPC succeeds to combine the implementations of QC and QNNs with the same hardware design of MPD as a uniquely valuable unconventional hardware architecture.

{\color{black} NLSE} solution is very important in the analysis and performance measurement of fiber optic cables \cite{wahls2015fast, tracy1988nonlinear}. It is also necessary for the solution of nonlinear Fourier transform (NLFT) which is a transformation that finds a wide range of applications {\color{black}with} increasing importance \cite{wahls2015fast}. NLSE and NLFT play a similar role for nonlinear and integration equations compared with the role of the FT in linear systems. NLFT is a transformation system for expressing the signal in the time plane by using nonlinear periodic waves or solitons  \cite{wahls2015fast}. It is also referred to as scattering transform. NLSE is expressed as follows \cite{wahls2015fast, tracy1988nonlinear}:
\begin{equation}
\imath \, \frac{\delta q(x, t)}{\delta t} \, + \, \frac{\delta^2 q(x,t)}{\delta x^2}  \, + \, 2 \, \kappa \, \vert q(x, t) \vert^2 \, q(x,t) = 0
\label{Eq_new_1}
\end{equation}
where $ q (x, t) $ is the solution wave function that provides the periodic boundary condition ($q (x + l, t) = q (x, t) $ and period $ l > 0 $) and $\kappa$ is some variable. In this article, the speed up in NLSE solution is conjectured by exploiting RTF summations  in QPC.

The remainder of the paper is organized as follows.  We firstly define and theoretically model photonic QPC and its extension for FO followed by the discussion of the performance based on Wigner distribution function. Then, future applications including QS, quantum neuron implementation and solution of NLSE are introduced, theoretically modeled  and the challenges are discussed. Numerical analysis for photonic MPD is provided and then open issues for realizing photonic QPC are presented.  

\section*{Results}

\subsection*{Quantum Path Computing with Optical Multi-plane Diffraction and Coherent Light Sources}
\label{sec01}
 
MPD set-up introduced in Ref. \citen{gulbahar2019quantumpath} as shown in Fig. \ref{Fig1} is extended to optical implementations for QC by using coherent laser sources and conventional photodetectors.  The set-up is composed of $N-1$ diffraction planes with $K_j$ slits on each plane for $j \in [1, N-1]$ and a single sensor plane indexed with $N$ while the {\color{black}central position} of {\color{black}a} slit  is given by $X_{j,i}$ for $i \in [1, K_j]$ as shown in Fig. \ref{Fig1}(a). Each slit is assumed to apply a spatial filtering of $ {\color{black} G(X_{j,i}, \widetilde{\beta}_{j,i}, x_j) \,\equiv  \,}\exp\big(-(x_j - X_{j,i})^2 \, / \, (2\, \widetilde{\beta}_{j,i}^2) \big)$, i.e., slit mask function, where $\widetilde{\beta}_{j, i}$ determines the slit width. The wave function on $j$th plane is denoted with $\Psi_j(x_j)$ which is the wave form after diffraction through the previous planes, i.e., with the indices $k \in [1, j -1]$, while before diffraction through the slits on  $j$th plane. There is an exponentially increasing number of propagation paths through the slits until to the final sensor plane, i.e., $N_p \equiv \prod_{j=1}^{N-1} K_j$, while $n$th path includes the diffraction through a single slit on each plane with the corresponding wave function $\Psi_{j, n}(x_j)$ on $j$th plane. Assume that $n$th path passes through the slit indexed with $s_{n,j}$ on $j$th plane and we define the path vectors $\vec{x}_{N-1, n}^T \equiv \begin{bmatrix} X_{1,s_{n,1}} \,\,X_{2,s_{n,2}} \,\, \hdots \,\,  X_{N-1,s_{n,N-1}} \end{bmatrix}$ and $\vec{x}_{N-1, \vec{s}}^T \equiv \begin{bmatrix} X_{1,s_{1}} \,\,X_{2,s_{2}} \,\, \hdots \,\,  X_{N-1,s_{N-1}} \end{bmatrix}$ where $\vec{s}  \equiv \begin{bmatrix}  s_{1}  \,\, s_{2}  \,\, \hdots \,\,   s_{N-1}  \end{bmatrix}$ and $(.)^T$ is the transpose operation. The mapping between the path index $n$ and slit index $s_{n,j}$ for the path is defined with the function $n = f_{s2n}(\vec{s})$ where $s_{n,j}$ is predefined for each $n$. Furthermore, $\vec{0}_{k}$  is the column vector  of length $k$ with all zeros and $\mathbf{0}_{k}$ is the square matrix of all zeros  with the size $k \times k$. Similarly, rectangular matrices are shown with $\mathbf{0}_{k,l}$. In the rest of the article, a parameter $B$ depending  only on $\vec{\beta}_{N-1, n} \equiv [\widetilde{\beta}_{1,s_{n,1}} \,\, ...\,\, \widetilde{\beta}_{N-1,s_{n,N-1}}]$ but not on $\vec{x}_{N-1, n}$ is denoted with $\widetilde{B}_{j, n}$ on each $j$th plane including  $\widetilde{.}$ over the symbol.  Therefore, if the slits are chosen with the same $\widetilde{\beta}_{j, {s_{n,j}}} = \beta_j$ specific to each plane, then $\widetilde{B}_{j, n}$ becomes independent of $n$ and is converted to the notation $B_{j}$.   
 
MPD set-up shown in Fig. \ref{Fig1} is utilized for QC denoted by QPC in Ref. \citen{gulbahar2019quantumpath} by sampling the interference of exponentially increasing number of interfering paths. The capability to theoretically characterize QPC with quantum FO provides future applications for both QC and  quantum information theory by exploiting energy efficient combination of optical elements.  Intensity sampling  on the sensor plane ($I[k]$) for the MPD set-up in Refs. \citen{gulbahar2019quantumpath, gulbahar2018quantum} generates a black-box (BB) function $f_{BB}[k]$ with the following special form  to utilize in solutions of important and classically hard number theoretical problems: 
 \begin{align}
 \label{Eq_new_2} 
\begin{split} 
  f_{BB}[k] \, \equiv  \,  I[k] \equiv \Bigg \vert \sum_{s_1 =1}^{K_1} \hdots \sum_{s_{N-1} =1}^{K_{N-1}}  \, e^{(\widetilde{A}_{\vec{s}} \,  + \, \imath \, \widetilde{B}_{\vec{s}}) (k \, T_s)^2}  \,\widetilde{\Upsilon}_{\vec{s}}    \,  e^{  \vec{x}_{N-1, \vec{s}}^T \, \mathbf{\widetilde{H}}_{\vec{s}}  \, \vec{x}_{N-1, \vec{s}}  }   
 \,   e^{(\vec{\widetilde{h}}_{\vec{s}}^T  \,  \vec{x}_{N-1, \vec{s}} )  \, k \, T_s}      \Bigg \vert^2  &    
\end{split}
\end{align}  
where  $k \in \mathbb{Z}$, $T_s \in \mathbb{R^+}$ is a sampling interval, $\widetilde{A}_{\vec{s}} \in \mathbb{R^-}$,  $\widetilde{B}_{\vec{s}} \in \mathbb{R^+}$ and  $\widetilde{\Upsilon}_{\vec{s}} \, \in \mathbb{C}$. The complex valued matrix   $\mathbf{\widetilde{H}}_{\vec{s}} \equiv \mathbf{\widetilde{H}}_{R,\vec{s}} \, + \, \imath\, \mathbf{\widetilde{H}}_{I, \vec{s}}$ and the vector $\vec{\widetilde{h}}_{\vec{s}} \equiv \vec{\widetilde{c}}_{\vec{s}} + \, \imath\,\vec{\widetilde{d}}_{\vec{s}}$  have the values depending on $\widetilde{\beta}_{j,i}$ for $j \in [1, N-1]$  and $i \in [1, K_j] $ corresponding to the specific selection of slits in the path  $\vec{s}$, inter-plane durations for the particle propagation, particle mass $m$ (for electron based set-ups in Refs. \citen{gulbahar2019quantumpath, gulbahar2018quantum}), beam width  $\sigma_0$ of the Gaussian source wave packet and Planck's constant $\hbar$.  The calculation of (\ref{Eq_new_2}) in  an  efficient manner is significantly hard while two different methods utilizing (\ref{Eq_new_2}) for practical problems are introduced. Solution for the partial sum of {\color{black}RTF}  or multi-dimensional theta function is the first application with importance in number theory and geometry   \cite{riemann1857theorie, deconinck2004computing, mumford2007tata, osborne2002nonlinear}.  The second method utilizes MPD with the phase term $\vec{\widetilde{d}}_{\vec{s}}^T  \,  \vec{x}_{N-1, \vec{s}}$ in $ \exp \big( \vec{\widetilde{h}}_{\vec{s}}^T  \,  \vec{x}_{N-1, \vec{s}}   \, k \, T_s \big)$  for period finding \cite{nielsen2002quantum} and the solution of specific instances of SDA problems \cite{lagarias1985computational}. 

{\color{black} The basic unit of QC systems, i.e., the qubit, is defined on a two-state system where discretized degrees of freedom (DoF) of photons including  path, transverse-spatial
modes and time/frequency bins are exploited to create high-dimensional entanglement \cite{qubit1}. For example, multi-slit structures are already utilized to define spatial qudits by projecting the wave function into the transverse position and momentum Hilbert spaces through slits and characterizing their properties using their propagation in free space \cite{qubit2, qubit3}. The qubit states in Ref. \citen{qubit3} are expressed  in the basis $\ket{l}, \ket{r}$ representing the  photons  passing through either the left or the right slit. However, entangled multiple photons, e.g., photon pair $A$ and $B$, are conventionally generated through spontaneous parametric down-conversion (SPDC) to realize multi-photon entangled state, e.g., $ \big( 1 \, / \, \sqrt{2}  \big) \, \big( \ket{l_A} \ket{r_B}  \, +  \,  \ket{r_A} \ket{l_B} \big)$. The fundamental difference of MPD based qudits from multi-photon slit based entangled spatial qudits is the utilization of the tensor product structure for \textit{each single photon} in \textit{time domain} rather than spatially among multiple photons obtained through SPDC \cite{gulbahar2019quantumpath}. The projection events through the slits of consecutive planes are freely entangled at two different time instants denoted as QPE with the detailed modeling presented in Ref. \citen{gulbahar2018quantum} based on \textit{consistent histories} and \textit{entangled histories} frameworks. The presented \textit{free entanglement in time domain} provides an important advantage exploiting directly the classical light sources and not requiring the difficult coupling of multiple photons. The concept of free entanglement is introduced for boson sampling exploiting  boson statistics of a number of indistinguishable bosons while they still require multiple photons, and generation and detection mechanisms for single photons \cite{aaronson2011computational}.

\begin{figure}[!t]
\centering
\includegraphics[width=3.5in]{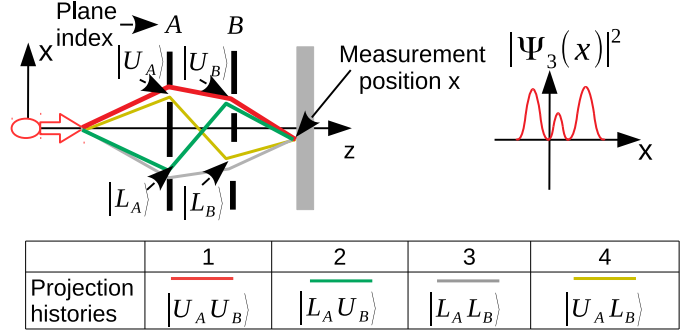}  
\caption{{\color{black}MPD based two-qubit state represented with four quantum histories of a single photon with the tensor product structure in time domain in analogy with the entangled state of two spatial qubits of two photons.}}
\label{Fig2}
\end{figure} 

History state in MPD is composed of diffraction events as follows \cite{gulbahar2018quantum}:
\begin{equation}
\label{eq0}
 \sum_{n} \pi_n \left[ \mathbf{P}_{N-1, s_{n,N-1}} \right]  \odot \left[ \mathbf{P}_{N-2, s_{n,N-2}} \right]  \odot ...  \odot \left[\mathbf{P}_{1, s_{n,1}} \right] \odot \left[ \rho_0 \right]
\end{equation}
where $\mathbf{P}_{j, s_{n,j}}$ represents the projection operator through the slit indexed with  $s_{n,j}$,  $\pi_n$ as $0$ or $1$ determines a compound set of trajectories, $\odot$ denotes tensor product operation and $ \left[ \rho_0\right]$ denotes the initial  state.  The analogy of MPD based multiple qubits with the general two qubit state of two photons is represented as shown in Fig. \ref{Fig2} in the basis of $\ket{U}$ and $\ket{L}$ for the upper and lower slits, respectively. The general state for the projection through two diffraction planes indexed with $A$ and $B$ is represented as follows:
\begin{equation}
\ket{\Psi_{{\color{black}3}}} \equiv a_{uu} \ket{U_A U_B} + \,  a_{lu} \ket{L_A U_B} + \,  a_{ll} \ket{L_A L_B} + \,  a_{ul} \ket{U_A L_B}
\end{equation}  
where the amplitudes are denoted by $a_{ij}$, and $i$ and $j$ denote the projection through upper or lower slits. The $A$ and $B$ in the MPD set-up denote the indices of planes for the projection of a single photon at different time instants rather than the indices of two photons as in the entangled state of two spatial qubits of two photons. There are four different projection history states where the wave function whose intensity to be measured on the final detection plane, i.e., $\Psi_{{\color{black}3}}(x)$ , is described as the interference of four different wave function histories corresponding to each trajectory, i.e., $\Psi_{{\color{black}3},j}(x)$ for $j \in [1,4]$:
\begin{equation} 
\label{eq1}
\Psi_{{\color{black}3}}(x) = \Psi_{{\color{black}3}, 1}(x) \, + \, \Psi_{{\color{black}3}, 2}(x) \, + \ \Psi_{{\color{black}3}, 3}(x) \, + \,  \Psi_{{\color{black}3}, 4}(x) 
\end{equation}
$\Psi_{{\color{black}3,}1}(x)$ corresponds to $a_{uu} \ket{U_A U_B}$ and the other components are defined as shown in Fig. \ref{Fig2}. Each component depends in a complex manner on the slit geometries as modeled by the {\color{black}RTF}.   QPC applications of MPD based high dimensional entangled states do not include any measurement by closing or opening slits but a final measurement on the detector plane obtaining the complicated interference pattern of exponentially many Feynman paths \cite{gulbahar2019quantumpath}.}
 
QPC based on {\color{black}FO} promises expanding the set of solvable problems both with LCT based general system design and also the sources including HG beams. Furthermore, a discussion is included to utilize non-Gaussian slits with the proposed mathematical modeling in the Open Issues and Discussion section. Propagation through Fourier optical systems based on Fresnel diffraction is modeled emphasizing the quantum nature of  Fresnel diffraction and FO in the Methods section.  Next, MPD modeling is proposed for Fresnel diffraction and arbitrary LCT based optical systems by utilizing the proposed kernels. 

\subsection*{Quantum Path Computing with Fourier Optical Systems}
\label{sec03}

\begin{figure}[!t]
\centering
\includegraphics[width=3.5in]{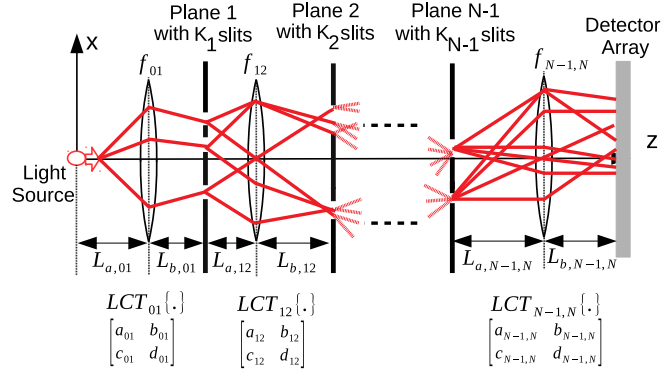}  
\caption{Photonic QPC architecture composed of classical light source, MPD set-up composed of $N-1$ diffraction planes with $K_j$ slits on $j$th plane, general LCT phase-space optics represented with the matrix elements $LCT_{j, j+1}$ between the planes indexed with $j$ and $j+1$, and a single sensor plane on which exponentially increasing number of propagation paths interfere. Each $LCT_{j, j+1}$ is implemented with sections of FSP for the lengths of $L_{a, j, j+1}$ and $L_{b, j, j+1}$, and a thin lens of focal length $f_{j, j+1}$ between them.}
\label{Fig3}
\end{figure} 

The set-up in Fig. \ref{Fig1} defined with FSP and electron based theoretical formulation is extended to  optical systems of LCT as shown in Fig. \ref{Fig3}. The kernel of one dimensional (1D) quadratic-phase system or LCT {\color{black} converting the input signal $f(x_0)$ to the output $\int_{-\infty}^{\infty} d x_0 \, K_{LCT}^{(a,b,c,d)}(x_1, x_0) \,f(x_0)$} is represented as 
follows:
\begin{equation}
\label{Eq_new_3}
K_{LCT}^{(a,b,c,d)}(x_1, x_0) \approx   \sqrt{\frac{1}{b}} e^{\frac{- \imath \, \pi}{4}} e^{  \frac{\imath \, \pi}{b}(d \, x_1^2 \, - 2 \, x_1 \, x_0 + \, a \, x_0^2)  }
\end{equation}
In matrix notation, it is shown with the following unit-determinant matrix:
\begin{equation}
\label{Eq_new_4}
\mathbf{M}_{LCT} = \begin{bmatrix}
a&b\\
c&d\\ 
\end{bmatrix}
\end{equation}
where $a \, d - b \, c =1$ and the matrix for the composition of two consecutive systems represented by $\mathbf{M}_1$ and $\mathbf{M}_2$ is calculated by the multiplication $\mathbf{M}_2 \, \mathbf{M}_1$. The kernel matrices $K_{FS}(x_1, x_0)$ and $K_{HO}(x_{1}, x_{0})$  denoting FSP kernel in phase-space optics \cite{ozaktas2001fractional} and the kernel based on quantum mechanical  harmonic oscillator (HO) modeling of the evolution of light wave function \cite{santos2018huygens} (in analogy with FRFT implementation),  respectively, are defined in the Methods section while discussing quantum FO.   Some simple examples of LCT matrices for propagation of length $L_{j, j+1}$ between $j$th and $(j+1)$th planes are given as follows \cite{ozaktas2001fractional}:
\begin{eqnarray}
\label{examp1}
\mbox{Free space propagation:  } \mathbf{M}_{FS}  & \equiv & \begin{bmatrix}
1& \frac{ 2 \, \pi \, \hbar \, L_{j, j+1} }{m_{\lambda} \, c } \\
0 &1\\
\end{bmatrix}  \\
\label{examp2}
\mbox{Fourier transform:  } \mathbf{M}_{FT}  & \equiv & \begin{bmatrix}
0 & 1\\
-1 & 0\\
\end{bmatrix}  \\
\label{examp3}
\mbox{Fractional Fourier transform of order $\alpha$:  } \mathbf{M}_{FRFT}  & \equiv & \begin{bmatrix}
\cos(\alpha) & \sin(\alpha)\\
-\sin(\alpha) & \cos(\alpha)\\
\end{bmatrix} \\
\label{examp4}
\mbox{Thin lens of focal length $f$:  } \mathbf{M}_{L}  & \equiv & \begin{bmatrix}
1 & 0\\
-\frac{1}{\lambda \, f} & 1\\
\end{bmatrix} \\
\label{examp5}
\mbox{Scaling:  } \mathbf{M}_{S}  & \equiv & \begin{bmatrix}
a_{j, j+1} & 0\\
0 & a_{j, j+1}^{-1}\\
\end{bmatrix} 
\end{eqnarray}
where $\mathbf{M}_{S}$ scales with  $\Psi_{j+1}(x_{j+1}) \equiv (1 \, / \, \sqrt{a_{j, j+1}}) \, \Psi_j(x_{j+1} \, / \, a_{j, j+1})$, {\color{black}  the kernel for $\mathbf{M}_{L}$ is $e^{  - \imath \, \pi \,  x_1^2  \, / \, (\lambda \, f)}  $} and $m_{\lambda} \equiv 2 \, \pi \, \hbar \,  /  \,  ( \lambda \, c) $ is defined in the Methods section after discussing (\ref{Eq_new_33}).

The varying forms of wave functions on the measurement plane extending (\ref{Eq_new_2}) are modeled which are promising to be utilized in QC applications. It is presented next such that obtained forms are similar to (\ref{Eq_new_2}) while having higher flexibility of system design. The theoretical modeling of BB functions for quantum HO based or FRFT based light propagation modeling with Gaussian sources is presented next with the wave function in (\ref{Eq_new_7}). Gaussian source case is also extended to arbitrary LCTs. Similarly, the wave functions for arbitrary LCTs with  HG sources are presented in (\ref{Eq_new_11})  and (\ref{Eq_new_13}) next. 

An arbitrary LCT with the matrix parameters $\lbrace a_{j, j+1}, b_{j, j+1}, c_{j, j+1}, d_{j, j+1} \rbrace$ is implemented in phase-space optics by consecutive applications of FSP of length $L_{a, j, j+1}$, then thin lens of focal length $f_{j, j+1}$, and another FSP of length  $L_{b, j, j+1}$ \cite{healy2015linear}. LCT matrix $\mathbf{M}_{LCT} $ is calculated as follows:
\begin{equation}
\label{Eq_new_5}
\mathbf{M}_{LCT} \equiv \begin{bmatrix}
1& \frac{ 2 \, \pi \, \hbar \,\,\tau_{b, j}^\star}{m_{\lambda} } \\
0 &1\\
\end{bmatrix} \, \begin{bmatrix}
1& 0 \\
-\frac{ 1}{ \lambda \, f_{j, j+1}}  &1\\
\end{bmatrix} \, \begin{bmatrix}
1& \frac{ 2 \, \pi \, \hbar \,\tau_{a, j}^\star}{m_{\lambda}} \\
0 &1\\
\end{bmatrix} 
\end{equation}
where $\tau^\star_{a, j} \equiv  L_{a, j, j+1} \, / \, c$ and $\tau^\star_{b, j} \equiv  L_{b, j, j+1} \, / \, c$, and the middle matrix is for the effect of thin lens \cite{ozaktas2001fractional}. FRFT with scaling is a special case of LCT as discussed in the Methods section. Therefore, FSP, FRFT and arbitrary LCT based QPC set-ups are implemented with the universal configuration in Fig. \ref{Fig3}.
 
\subsubsection*{QPC with Fresnel Diffraction and FRFT by using Gaussian Sources}
\label{sec0301}

{\color{black} Firstly, two special cases of LCTs are considered, i.e.,  FSP of light and propagation modeled with FRFTs denoting graded-index media propagation as the solution of the quantum HO in Ref. \citen{santos2018huygens}.} Furthermore, we assume that the source distribution has a Gaussian form of $ \Psi_0(x_0) = \mbox{exp}\big(- \, x_0^2 \, / \, (2 \, \sigma_0^2) \big)  \, /   \, \sqrt{\sigma_0 \,\sqrt{\pi }} $ while HG waveforms as eigenfunctions of FRFTs  \cite{ozaktas2001fractional} are considered for the general case of LCTs in the next section. It is assumed that the optical system between the planes results in the kernels {\color{black} $K_{FS}(x_1, x_0)$ and $K_{HO}(x_1, x_0) $} defined in (\ref{Eq_new_31})  and (\ref{Eq_new_33}) based on Fresnel diffraction integral for free space and {\color{black}quantum} HO solution \cite{santos2018huygens}, respectively. {\color{black}The definition and the derivation of HO based kernel with the following kernel matrix for the propagation duration of $t_{01}$ are detailed in the Methods section while we are discussing the quantum mechanical modeling of FO, i.e., denoting with quantum FO:
\begin{equation}
\mathbf{M}_{HO}   =   \begin{bmatrix}
\cos(\omega \, t)& \frac{ 2 \, \pi \, \hbar \,t_{01}\, \sin(\omega \, t)}{m_{\lambda}}\\
-\frac{ m_{\lambda} \, \sin(\omega \, t)}{2 \, \pi \,  \hbar \,t_{01}} &\cos(\omega \, t)\\
\end{bmatrix}
\end{equation}
The important observation is that} iterative integration  with $K_{HO}(x_1, x_0)$  results in the final intensity distribution of MPD with the same form of $K_{FS}(x_1, x_0)$ while  with different {\color{black} algorithms for calculating the} iteration parameters as shown in Table \ref{Table_Gaussian} in the Methods section. The kernel $K_{FS}(x_1, x_0)$ has the same form with $K_{m, FS}(x_1, x_0)$ used for QPC modeling in Ref. \citen{gulbahar2019quantumpath} by replacing the electron mass $m$ with the photon equivalent mass $m_{\lambda}$. Therefore, the same formulations are utilized {\color{black}for the cases of FS and HO solutions while modeling the {\color{black}sampled} wave function on the sensor plane with iterations and the resulting structure of problem solving capabilities.}  

The wave function for $n$th path on the sensor plane for the general case of non-uniform slit widths is given by the following by using the  iterative formulation:
\begin{align} 
\begin{split}
\label{Eq_new_6}
\Psi_{N,n}^G (x_{N})  \, = \, & \chi_{0} \, \bigg( \prod_{j=1}^{N-1} \chi_{j,n} \bigg) \, e^{(\widetilde{A}_{N-1, n} \,+ \, \imath \, \widetilde{B}_{N-1, n})\, x_{N}^2}    \,e^{(C_{N-1,n}  \,+ \, \imath \, D_{N-1,n})\, x_{N} }   
\end{split}
\end{align} 
It is further simplified by extraction of $\vec{x}_{N-1, n}$ dependent parts and summing the contributions from each path as follows:
\begin{align}
\label{Eq_new_7}
\begin{split}
\Psi_{N}^G(x_{N})  \,\equiv\,  {\color{black} \sum_{n=0}^{N_p-1} \Psi_{N,n}^G (x_{N}) \, =  \, } \sum_{n=0}^{N_p-1}  \widetilde{\Upsilon}_{N, n}^G  \, e^{ \vec{x}_{N-1,n}^T \, \mathbf{\widetilde{H}}_{N-1, n}^{HO/G}  \, \vec{x}_{N-1,n} }   \, e^{(\widetilde{A}_{N-1,n} \, + \,  \imath \, \widetilde{B}_{N-1,n} ) \, x_N^2}    
 \, e^{(\vec{\widetilde{h}}_{N-1,n}^T   \vec{x}_{N-1,n}) \, x_N} &       
\end{split} 
\end{align}
where $\widetilde{\Upsilon}_{N, n}^G  \equiv \chi_0 \, \big  ( \prod_{j=1}^{N-1} \sqrt{\widetilde{\xi}_{j, n}}  \big  )$, and the complex vector $\vec{ \widetilde{h}}_{N-1,n}$ and the matrix $\mathbf{\widetilde{H}}_{N-1, n}^{HO/G}$ are defined in the Methods section {\color{black} for the HO case with simplified formulation} compared with the  {\color{black}case} for electron based FSP set-up in Ref. \citen{gulbahar2019quantumpath}. The corresponding iteration parameters are given in Table \ref{Table_Gaussian} in the Methods section.
 
We have not included the effects of special forms of $K_{HO}(x_{j+1}, x_j)$ with $\omega \, t_{j, j+1}  \, = \, k \, \pi$  for $k \in \mathbb{Z}$  {\color{black}corresponding to integer multiples of FRFT order $2$} since the result is $\Psi_{j+1}(x_{j+1}) \equiv \Psi_j(\pm x_{j+1})$ (inserting $\pm x_{j+1}$ into $\Psi_j(x_{j})$ ) \cite{ozaktas2001fractional}. This case can be simply realized by assuming that spatial filtering operations of the slits on $j$th and $(j+1)$th planes are combined on a single plane by also noting that whether the wave function is reversed or not. For example, multiple inter-plane propagation intervals can result in multiple reversals with the overall effect of the identity and combined spatial filtering of Gaussian slits.  

\subsubsection*{QPC with Arbitrary Linear Canonical Transforming Optical Systems}
\label{sec0302}

\textbf{Gaussian Sources:} The resulting final intensity of MPD propagation for the case of $K_{LCT}^{(a,b,c,d)}$  ($b_{j, j+1} \neq 0$) with  Gaussian sources  has the same form with $K_{HO}$  while with different {\color{black}algorithms for calculating} iteration parameters in Table \ref{Table_Gaussian} in the Methods section and replacing $\widetilde{\mathbf{H}}_{N-1, n}^{HO/G}$ with $\widetilde{\mathbf{H}}_{N-1, n}^{LCT/G}$.  Therefore, all the derivations utilized for $K_{HO}$ including the explicit forms of the wave function are applicable.  We have not included $K_{LCT}^{(a_{j, j+1},\, b_{j, j+1},\, c_{j, j+1},\, d_{j, j+1})}$ with $b_{j, j+1} = 0$  for simplicity. Two simple cases are scaling  and  chirp multiplication with $a_{j, j+1} = d_{j, j+1} = 1$  resulting in  $\Psi_{j+1}(x_{j+1}) \, \equiv \, \exp( \imath \, \pi \, c \, x_{j+1}^2) \, \Psi_j(x_{j+1})$ \cite{ozaktas2001fractional}. These cases further improve the flexibility of the LCT system to realize the desired transformation on the wave function. 

\begin{figure*}[!t]
\centering
\includegraphics[width=6.5in]{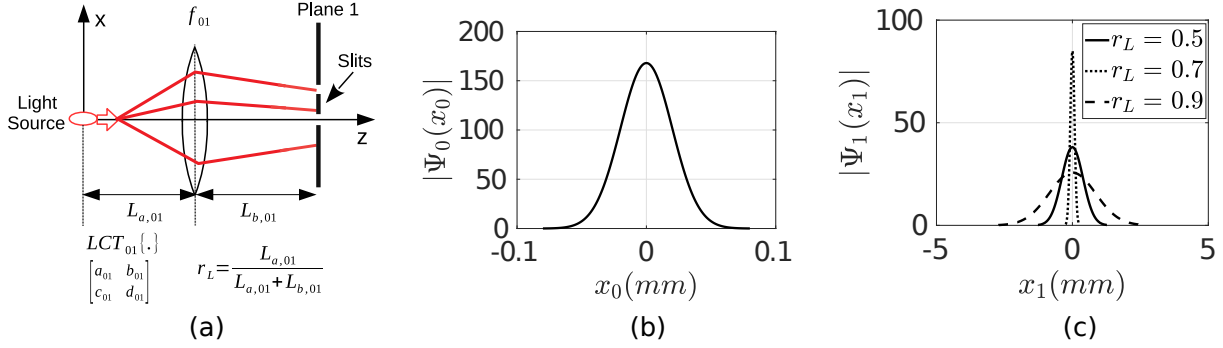}
\caption{(a) The set-up with a single thin-lens of focal length $60$ (mm), (b) Gaussian source with $\sigma_0 = 20$ ($\mu$m) and $\lambda = 650$ (nm), and (c) the distribution of the wave function on the first plane by shifting the lens inside the spatial interval of fixed total length $L_{01} = t_{01} \times c$ with $t_{01} = 1$ (ns) by varying  the ratio  of $r_{L} = L_{a, 01} \, / \, (L_{a, 01} + L_{b, 01})$.}
\label{Fig4}
\end{figure*} 

A simple example is presented with explicit expressions as follows for $K_1 =3$, $K_2 = 2$, $N = 3$, $b_{j,j+1}  \neq 0$ for $j \in [0,2]$, fixed slit width parameters $\beta_1$ and $\beta_2$ for simplicity and $\vec{x}_{2, n}^T \equiv \begin{bmatrix} X_{1,s_{n,1}} \,\,X_{2,s_{n,2}} \end{bmatrix}$:
\begin{align}
\label{Eq_example0}
\begin{split}
\Psi_{3}^G(x_{3}) =  \sum_{n=0}^{5}   \Upsilon_{3}^G  \, e^{ \vec{x}_{2,n}^T \, \mathbf{H}_{2}^{LCT/G}  \, \vec{x}_{2,n} }   \, e^{(A_{2} \, + \,  \imath \, B_{2} ) \, x_3^2}    
 \, e^{(\vec{h}_{2}^T   \vec{x}_{2,n}) \, x_3} &       
\end{split} 
\end{align}
where the following parameters are derived by using Table \ref{Table_Gaussian} in the Methods section:
\begin{align}
\label{Eq_example1}
\begin{split}
 \mathbf{H}_{2}^{LCT/G} =  \begin{bmatrix}
\dfrac{pol_{1}(\beta_2, \sigma_0)}{pol_{2}(\beta_1, \beta_2, \sigma_0)} & 0 \vspace{0.1in} \\
\dfrac{pol_{3}(\sigma_0)}{\imath \, pol_{2}(\beta_1, \beta_2, \sigma_0)} & \dfrac{pol_{4}(\beta_1, \sigma_0)}{pol_{2}(\beta_1, \beta_2, \sigma_0)}\\
\end{bmatrix} \,   &       
\end{split}  
\end{align}
\begin{align}
\label{Eq_example2}
\begin{split}
 \vec{h}_{2}^T =  \begin{bmatrix}
\dfrac{pol_{5}(\beta_1, \beta_2, \sigma_0)}{pol_{6}(\beta_1, \beta_2, \sigma_0)} & \dfrac{pol_{7}(\beta_1, \beta_2, \sigma_0)}{pol_{6}(\beta_1, \beta_2, \sigma_0)} \\
\end{bmatrix} + \imath  \, \begin{bmatrix}
\dfrac{pol_{8}(\beta_1, \beta_2, \sigma_0)}{pol_{6}(\beta_1, \beta_2, \sigma_0)} & \dfrac{pol_{9}(\beta_1, \beta_2, \sigma_0)}{pol_{6}(\beta_1, \beta_2, \sigma_0)} \\
\end{bmatrix}   &       
\end{split} 
\end{align}
\begin{align}
\label{Eq_example3}
\begin{split}
  \Upsilon_{3}^G =  -2 \, (-1)^{3/4} \, \sqrt{2} \,\pi^{5/4} \sqrt{\dfrac{pol_{10}(\beta_1, \beta_2, \sigma_0)}{pol_{11}(\beta_1, \beta_2, \sigma_0)}}
\end{split} 
\end{align}
\begin{align}
\label{Eq_example4}
\begin{split}
  A_2  \, + \,\imath  \,  B_2 =   \dfrac{pol_{12}(\beta_1, \beta_2, \sigma_0)}{pol_{13}(\beta_1, \beta_2, \sigma_0)} \, + \, \imath  \, \dfrac{pol_{14}(\beta_1, \beta_2, \sigma_0)}{ b_{23} \, pol_{13}(\beta_1, \beta_2, \sigma_0)}
\end{split} 
\end{align}
where  it is observed that the parameters are some rational complex polynomials of $\beta_1$, $\beta_2$ and $\sigma_0$ with the coefficients in terms of polynomial expressions of $a_{j, j+1}$, $b_{j,j+1}$ and $d_{j, j+1}$ for $j \in [0,2]$ with varying orders of ($\beta_1$, $\beta_2$, $\sigma_0$): reaching ($0$, $2$, $2$) for $pol_{1}$;  ($2 $, $ 2$, $ 2$) for $pol_{2}$; ($ 0$, $ 0$, $ 2$) for $pol_{3}$; ($2 $, $0 $, $ 2$) for $pol_{4}$; ($6 $, $ 4$, $8 $) for $pol_{5}$, $pol_{7}$, $pol_{8}$  and $pol_{9}$; ($8 $, $ 4$, $ 8$) for $pol_{6}$,  ($ 4$, $2 $, $ 5$) for $pol_{10}$; ($ 4$, $ 2$, $ 6$) for $pol_{11}$; ($4 $, $ 4$, $4 $) for $pol_{12}$,  $pol_{13}$ and  $pol_{14}$. 
 
The highly complicated expressions for the polynomials are explicitly shown in Table \ref{Table_Example_Gaussian} in the Methods section such that  they are obtained by using the iteration method in Table \ref{Table_Gaussian} in the same section.  It is possible by using the explicit expressions directly to perform various gedanken experiments and computational complexity analysis with any number of slits and LCTs.
 
A simple numerical example is presented as shown in Fig. \ref{Fig4}. The scaling property of thin lens is utilized in the Numerical Results section to improve the intensity of the diffraction on the final detection plane. For example, a simple Gaussian source beamwidth of $\sigma_0 = 20$ ($\mu$m) and $\lambda = 650$ (nm) shown in Fig. \ref{Fig4}(b) is scaled by shifting the position of the lens of focal length $60$ (mm) inside the interval of fixed length of $L_{01} = L_{a, 01} + L_{b, 01} = t_{01} \times c $ with $t_{01} = 1$ (ns) as shown in Fig. \ref{Fig4}(a). The shift is modeled with the ratio $r_{L} = L_{a,01} \, / \,  L_{01} $. It is observed in Fig. \ref{Fig4}(c) that the intensity of the wave function can be focused with respect to the positions of the slits on the first plane.

\noindent \textbf{Hermite-Gaussian Sources:} If the source is chosen as the standard HG waveform of $\Psi_0(x_0) \, = \, (2^{1/4} \, / \, \sqrt{W_0 \, 2^l \, l! }) \, \mbox{exp}\big(- \, \pi \, x_0^2 \, / \, W_0^2 \big) \, H_l( \sqrt{2 \, \pi} \, x_0 \, / \, W_0) $     for $K_{LCT}^{(a_{j, j+1},\, b_{j, j+1},\, c_{j, j+1},\, d_{j, j+1})}$ ($b_{j, j+1} \neq 0$) where $H_l(x) \, \equiv \, (-1)^l \, e^{x^2} \, d^l e^{-x^2}\, / \,dx^l$ is the $l$th order Hermite polynomial \cite{ozaktas2001fractional, bateman1954tables},  then $\Psi_{N, n}(x_N)$ for $n$th path is obtained as follows by using the integral equality of HG functions in the Methods section:
\begin{align}  
\begin{split} 
\label{Eq_new_8}
\Psi_{N,n}^{HG}  (x_{N})  \, = \,  \chi_{01} \bigg( \prod_{j = 1}^{N-1} \chi_{j,j+1, n}  \bigg) \, e^{ \widetilde{u}_{N-1, N, n}\, x_{N}^2}   \, e^{\,v_{N-1, N, n} \, x_{N} }  \, H_{l}(\widetilde{g}_{N-1, N, n} \, x_{N} \, + \, h_{N-1, N, n})  & 
\end{split}
\end{align}
where the parameters $\chi_{01}$, $\chi_{j,j+1, n}$, $\widetilde{u}_{j, j+1, n}$, $v_{j, j+1, n}$, $\widetilde{g}_{j, j+1, n}$ and $h_{j, j+1, n}$ obtained in an iterative manner  for  $j \in [1, N-1]$ are calculated with simple algebra for $n$th path and shown in Table \ref{Table_HG}  in the Methods section.  Simple algebraic manipulations of (\ref{Eq_new_8}) to extract $\vec{x}_{N-1,n}$ dependent parts result in the following simplification:  
\begin{align}
\begin{split}
\label{Eq_new_9}
\Psi_{N,n}^{HG}  (x_{N})  \, = \,  \chi_{01} \, \bigg( \prod_{j = 1}^{N-1} \widetilde{\chi}_{a, j,j+1, n}  \bigg) \, e^{\frac{- \, \imath \, \pi \, (N-2)}{4}}  \, e^{\vec{x}_{N-1, n}^T \, \widetilde{\mathbf{H}}_{N-1, n}^{LCT/HG} \, \vec{x}_{N-1, n} } \hspace{0.85in} & \\ 
\times \,  e^{ (\vec{\widetilde{\gamma}}_{N-1, n}^T \, \vec{x}_{N-1, n}) \,x_N }   \, e^{\widetilde{u}_{N-1, N, n} \,x_N^2} \, H_l(\widetilde{g}_{N-1, N, n} \, x_N  \, + \, \vec{\widetilde{\eta}}^T_{N-1, n} \,\vec{x}_{N-1, n}  ) &
\end{split}
\end{align} 
where  $\widetilde{\chi}_{a, j,j+1, n}$ for $j \in [1, N-1]$ is defined in Table \ref{Table_HG},  the vectors $ \vec{\widetilde{\gamma}}_{N-1, n}$ and $\vec{\widetilde{\eta}}_{N-1, n}$,  and the matrix $\widetilde{\mathbf{H}}_{N-1, n}^{LCT/HG}$ are defined in the Methods section.  It is observed in (\ref{Eq_new_9}) that each different path results in a different shift on Hermite polynomial determined with $\vec{\widetilde{\eta}}^T_{N-1, n} \,\vec{x}_{N-1, n} $ even for the uniform $\beta_j$ for each slit on $j$th plane. As a result, the final wave function on the sensor plane denoted with $\Psi_{N}^{HG}  (x_{N})$ for the general case of non-uniform slit widths defined with $\widetilde{\beta}_{j, n}$ for $j \in [1, N-1]$ and $n \in [0, N_p -1]$ is given by the following:
\begin{eqnarray} 
\label{Eq_new_10}
\Psi_{N}^{HG}  (x_{N})  \,  = &&  \, \sum_{n = 0}^{N_p -1} \Psi_{N,n}^{HG}  (x_{N})  \hspace{1.5in}    \\
\label{Eq_new_11}
  \,   =  && \, \sum_{n = 0}^{N_p -1}  \widetilde{\Upsilon}_{N, n}^{HG}  \,  e^{\vec{x}_{N-1, n}^T \, \widetilde{\mathbf{H}}_{N-1, n}^{LCT/HG} \, \vec{x}_{N-1, n} }   \, e^{\widetilde{u}_{N-1, N, n} \,x_N^2} \,   e^{ (\vec{\widetilde{\gamma}}_{N-1, n}^T \, \vec{x}_{N-1, n}) \,x_N }   \,    H_l(\widetilde{g}_{N-1, N, n} \, x_N  \, + \, \vec{\widetilde{\eta}}^T_{N-1, n} \,\vec{x}_{N-1, n}  )    
\end{eqnarray}  
where $\widetilde{\Upsilon}_{N, n}^{HG}   \, \equiv \, \chi_{01} \, \big( \prod_{j = 1}^{N-1}  \widetilde{\chi}_{a, j,j+1, n}  \big) $ $ e^{ - \, \imath \, \pi \, (N-2) \, / \,4} $ and {\color{black}with the} similarity to the form in (\ref{Eq_new_7}) {\color{black} for the HO solution} except multiplicative Hermite polynomial for each $n$th path. The complexity of calculating the  Gaussian form in (\ref{Eq_new_7}) is classically hard as thoroughly discussed in Ref. \citen{gulbahar2019quantumpath} which requires to compute a special form of partial sum of {\color{black}RTF} while the complex vector $\vec{ \widetilde{h}}_{N-1,n}$ and the matrix $\mathbf{\widetilde{H}}_{N-1, n}^{HO/G}$ varying for each path making it much harder compared with the computation of conventional partial sum of {\color{black}RTF}. Therefore, 
the complexity characterization of computing $\Psi_{N}^{HG}  (x_{N}) $  is an open issue while it is expected to be significantly hard since each summation term depends on path index $n$ with varying vector and matrix parameters while also including a product term of Hermite polynomial for each path making it harder. 
 
If the uniform slit width case is chosen and the path independent variables are denoted with $\widetilde{\Upsilon}_{N, n}^{HG} = \Upsilon_{N}^{HG}$, $\widetilde{\chi}_{a, j,j+1, n} = \chi_{a, j,j+1}$,   $\widetilde{\mathbf{H}}_{N-1, n}^{LCT/HG} = \mathbf{H}_{N-1}^{LCT/HG}$,   $\vec{\widetilde{\gamma}}_{N-1, n}  = \vec{\gamma}_{N-1}$,   $\widetilde{u}_{N-1, N, n} = u_{N-1, N}$,  $\widetilde{g}_{N-1, N, n} = g_{N-1, N}$,  $\vec{\widetilde{\eta}}_{N-1, n} = \vec{\eta}_{N-1}$, then  (\ref{Eq_new_11}) is transformed into the following:
 \begin{eqnarray}
\label{Eq_new_12} 
 \Upsilon_{N}^{HG} \, e^{u_{N-1, N} \,x_N^2} \, \sum_{n = 0}^{N_p -1}  \,    e^{\vec{x}_{N-1, n}^T \, \mathbf{H}_{N-1}^{LCT/HG} \, \vec{x}_{N-1, n} }    \,  \, e^{ (\vec{\gamma}_{N-1}^T \, \vec{x}_{N-1, n}) \,x_N } \, H_l(g_{N-1, N} \, x_N  \, + \, \vec{\eta}^T_{N-1} \,\vec{x}_{N-1, n}  )     &&
\end{eqnarray}
It is further simplified  as follows by using the useful identity $H_{l}(x \, + \, y) = (H +  2\,y)^l$ in Ref. \citen{weisstein2002hermite} where $H^k \equiv H_k(x)$:
\begin{align}
\begin{split}
\label{Eq_new_13} 
\Psi_{N}^{HG, U}  (x_{N})  \,  =  \,   \Upsilon_{{\color{black}N}}^{HG} \, e^{ u_{N-1, N} \,x_N^2} \,  \sum_{n = 0}^{N_p -1}  \,     e^{ (\vec{\gamma}_{N-1}^T \, \vec{x}_{N-1, n}) \,x_N } \,  e^{\vec{x}_{N-1, n}^T \, \mathbf{H}_{N-1}^{LCT/HG} \, \vec{x}_{N-1, n} } \, (H^\star(x_N) \, + 2 \, \vec{\eta}^T_{N-1} \,\vec{x}_{N-1, n})^l      &  
\end{split}
\end{align} 
where   $\big(H^\star(x_N)\big)^k \, \equiv \, H_k(g_{N-1, N} \, x_N)$. The computational complexity of calculating  $\Psi_{N}^{HG, U}  (x_{N}) $ is similarly expected to be significantly hard since the mathematical form is more complicated compared with the partial sum of {\color{black}RTF}.

The set-up parameters including the slits, lenses and inter-plane distances are required to be tuned in order to obtain the desired vectors   $\vec{\widetilde{h}}_{N-1,n}$, $\vec{\widetilde{\gamma}}_{N-1,n}$, $\vec{\widetilde{\eta}}_{N-1,n}$ and matrices $\widetilde{\mathbf{H}}_{N-1, n}^{HO/G}$, $\widetilde{\mathbf{H}}_{N-1, n}^{LCT/G}$ and $\widetilde{\mathbf{H}}_{N-1, n}^{LCT/HG}$  in (\ref{Eq_new_7}), (\ref{Eq_new_11}) and (\ref{Eq_new_13}) for the targeted number theoretical problems.  Next,  Wigner distribution is defined where its  negative volume is regarded as an indicator  of non-classicality.

\subsection*{Wigner Distribution, Negativity and Path Magnitudes}
\label{sec04}

The momentum domain wave function  $\Psi_{p,j}(p_j)$ is defined as  Fourier transform of spatial representation of wave function $\Psi_j(x_j)$ on $j$th plane as follows:
\begin{equation}
\label{Eq_new_14} 
\Psi_{p, j}(p_j) = \frac{1}{\sqrt{2 \, \pi \, \hbar}} \int d x_j \, \Psi_j(x_j) \exp \big( - \imath \, x_j \, p_{\color{black}j} \, / \, \hbar\big) 
\end{equation}
The distribution of energy through space-momentum phase-space is described by Wigner distribution function defined as follows \cite{kenfack2004negativity, gulbahar2019quantumpath}:
\begin{equation}
\label{Eq_new_15} 
W_{j}(x_j, p_j) = \frac{1}{\pi \, \hbar} \int dy \, \Psi_j(x_j \,- \, y) \,  \Psi_j^*(x_j \,+ \, y) \, e^{  \frac{ \imath \, 2 \, p_j \,  y}{\hbar} }
\end{equation} 
The negative volume of Wigner function defined in Ref. \citen{kenfack2004negativity} and utilized in Ref. \citen{gulbahar2019quantumpath} to describe the increasing non-classicality or time-domain entanglement resources in Ref. \citen{gulbahar2018quantum} is described as $V_j   \equiv$ $ \big(\int \int \vert W_j(x_j, p_j)\vert \, d x_j \, d p_j  \, -  \, 1 \big) \, / \, 2$. On the other hand, the probability of the particle to be detected on $j$th plane, i.e., to be diffracted through $(j-1)$th plane, is computed as follows:
\begin{equation}
\label{Eq_new_16} 
P_E(j) \equiv \int d x_j \, \vert \Psi_j(x_j)\vert^2
\end{equation}
In this article, the increasing interference complexity and non-classical nature of MPD based time-domain entanglement resources are {\color{black}assumed to be} characterized by utilizing $V_j$ and by observing the magnitudes of the interfering paths defined as $P_{E,n}(j) \equiv \int d x_j \, \vert \Psi_{j, n} (x_j)\vert^2$ for each $n$th path. Therefore, more paths with large magnitudes and $V_j$ emphasize  increasing interference and non-classicality. Characterizing the correlation between the distribution of path magnitudes and $V_j$ is an open issue since the behavior is highly set-up specific as observed in the Numerical Results section. On the other hand, path magnitudes throughout the whole plane may not reflect their localized characteristics such as effecting some sample locations more compared with the others. Therefore, it is an open issue to characterize the interference complexity in terms of the intensity distribution of the paths while the path magnitudes are taken as a reference for simplicity in this article.   Next, the potential future applications of QPC architecture based on {\color{black}FO} are presented.
 
\subsection*{Future Applications}

\subsubsection*{Applications for Quantum Supremacy and Certified Random Number Generation}

\begin{figure*}[!t]
\centering
\includegraphics[width=6.5in]{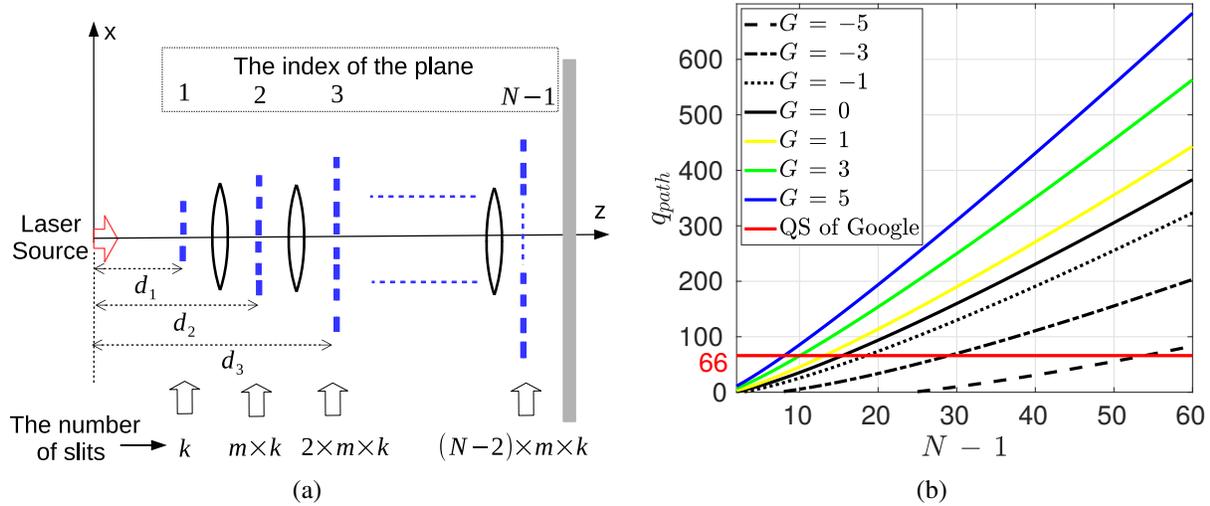}
\caption{(a) The special QPC design with N-1 planes diffracts the quantum wave function expanding with planes consisting of a linearly increasing number of slits. The positions of the planes ($d_j$), the number of slits ($k$) in the first plane and the linear increment ratio $m$, i.e., the number of slits growing with $j \times m \times k$, allow flexible design according to the width of the laser beam. (b) The virtual qubit Hilbert space size corresponding to the number of Feynman paths versus the number of planes for varying gain variable ($G$). QS experiments of Google have recently been performed with $ \approx 2^{66}$ Feynman paths  \cite{arute2019quantum}. Even at significant loss rates of the QPC system, thanks to dozens of planes, the Hilbert space size reaches hundreds of virtual qubit levels and achieves very strong QS capability. It is promising as an alternative system design for next generation QS experiments.}
\label{Fig5}
\end{figure*} 
  
The scalable structure of QPC with  coherent light sources and low complexity {\color{black}FO} promise the large scale implementation of QS experiments. Recently, some experiments are performed with 53 qubits with superconductor based architectures of Google as a milestone for the human history of computational capabilities \cite{arute2019quantum, cho2019google}.  {\color{black}In this article, we propose that a similar experiment} could be formulated in the QPC set-up as the problem of finding the distribution of light intensity on the photodetector array plane by randomly generated slit positions and widths analogical to random circuit sampling \cite{bouland2019complexity, bouland2018quantum}. Fig. \ref{Fig5}(a) shows the system architecture that is similar to the randomly generated quantum gates. The aim is to perform a complexity analysis of a randomly generated QPC architecture. The total number of Feynman paths is expressed as follows {\color{black}for $L$ diffraction planes}:
\begin{eqnarray}
N_{path} = && \bigg( \prod_{i =1}^{L {\color{black}-1}} i \bigg) \, m^{L-1} \, k^L 
\label{Eq_new_17} 
\end{eqnarray}
Suppose that energy decreases by $1/s$  ($s > 1$) to a total of $1/s^L$ and the number of significant paths decreases by  $1/r$ ($r > 1$) leading to a decrease in the total number of effective paths by  $1/r^L$ approximating the final intensity distribution. Therefore, if we define the number of paths that can be realized for unit {\color{black}source} energy with  $\widetilde{N}_{path}$, then the following is obtained:
\begin{eqnarray}
\label{Eq_new_18} 
\widetilde{N}_{path} = && N_{path} \,  \frac{1}{s^L} \, \frac{1}{r^L} \\
\label{Eq_new_19} 
  = &&\bigg( \prod_{i =1}^{L{\color{black}-1}} i \bigg) \, m^{L-1} \, k^L  \, \frac{1}{s^L} \, \frac{1}{r^L}
\end{eqnarray}
If we measure the Hilbert space size created by the total number of paths by defining the virtual qubit number and assuming $s \equiv 2^{s^*}$,  $r \equiv 2^{r^*}$, $m \equiv 2^{m^*} {\color{black}2^2}$ {\color{black}(assuming the minimum of $m = 8$ with $m^* = 1$)},  $k \equiv 2^{k^*}$, then the following definition is obtained:
\begin{eqnarray}
\label{Eq_new_20} 
q_{path} \equiv && \log_2 \big( \widetilde{N}_{path} \big) \\
\label{Eq_new_21} 
   =  && \log_2 \big( (L{\color{black}-1})! \big) \, - \, 2 \, - m^* + L\big(2 +  m^* + k^* - s^* - r^*\big)
\end{eqnarray}
 
The basic expression determining the size of the Hilbert space is denoted with the gain $G = m^* + k^* - s^* - r^*$. It provides the cumulative effect of increasing number of paths due to the {\color{black} linearly increasing number  of the slits with the coefficient $m$} and the initial number of the slits $k$ combined with the decreasing number of paths due to the inter-plane attenuation coefficients $s$  and $r$ for the effective number of the paths. As shown in Figure \ref{Fig5}(b) {\color{black}for $ L = N-1$}, even at very low gain rates, {\color{black} e.g.,  $G = -5$   with $m^* = 1$ }, that is, where the spreading energy drops very quickly and the number of significant paths is too low, the number of virtual qubits reaches hundreds. Furthermore,  $m^* + k^*$, which can be designed flexibly {\color{black}in a multi-slit architecture}, is adapted against the low gain. In addition, even with $N =10$ planes, Hilbert space size of {\color{black}approximately reaching} hundreds of virtual qubits is obtained for the case of high gain {\color{black}$G$}.  For example, assume the worst case situation such that diffracted photon forms a large amplitude path by diffracting through a locally limited number of slits on the next plane denoted by the parameter $\widetilde{r}$. In other words, the slit locations distant apart on consecutive planes will not form a large amplitude path and the number of effective paths increases as the multiplication by $\widetilde{r}^{j}$ after diffracting through $j$ consecutive planes (by assuming there is enough number of closely spaced slits on the next plane). This can be adjusted by increasing the inter-plane distance such that each diffracted beam will expand to a larger area on the consecutive plane. Moreover, assume that the slits are placed close enough to keep the probability of the diffraction through the next plane roughly constant, i.e.,  $1 \,/ \, s $, as observed in MPD simulation studies in Ref. \citen{gulbahar2019quantumpath}.  Then, if the condition $s < \widetilde{r}$ is satisfied, the number of effective paths will increase with the multiplication by $(\widetilde{r} \, / \, s)^{j}$  through $j$ consecutive planes. For example, effective number of paths in Ref. \citen{gulbahar2019quantumpath} is observed to be increasing with $(\widetilde{r} \, / \, s)^{3} > 70 $ even by removing many effective paths (Fig. 7(b) in Ref. \citen{gulbahar2019quantumpath}). In other words, assuming $ \widetilde{r} \, / \, s  \approx 2^2$ allows to reach $2^{100}$ effective Feynman paths with $ N \approx 51$ planes. Therefore, without requiring extensive simulations, it is clear that the number of paths increases exponentially with specially adjusted set-up parameters of inter-plane distance, slit widths and distributions.

Each path will form a unique contribution to the overall intensity. There is no apparent way of calculating the exact final intensity other than identifying and summing the contribution of each path. Then, increasing the complexity of MPD set-up makes it harder to calculate the contribution of each path until reaching to the QS scale.   It is possible to compare roughly with Google QS experiment where the computational complexity requires the calculation of $4^{31} \times 2^4  = 2^{66}$ different Feynman paths for $53$ qubits and $20$ cycles (Table XI in Ref. \citen{arute2019quantum}) as shown in Fig. \ref{Fig5}(b). Compared to the $q_{path} = 66$, the proposed QPC system architecture {\color{black}suggests to perform} future QS experiments with a simple system structure.
 
The open issues include the rigorous characterization of the computational complexity of sampling from MPD exploring the relations among the paths in terms of magnitude and distribution. Furthermore, the inter-plane gain $G$ is required to be both theoretically modeled and experimentally measured for random and large diffraction architectures.  Another open issue is to {\color{black}analyze the modeling of} the sampling problem of QPC with universal quantum circuits {\color{black}and} to determine the computational complexity {\color{black}class, e.g., the relations with $BQP$ \cite{aaronson2011computational} and complexity theoretical fundamentals of QS experiments \cite{aaronson2016}}. Experimental implementation requires slit design and manufacturing, sensitive photon detection due to the attenuation after large number of planes and spatially coherent light sources covering all the paths reaching to the detector plane. The number of slits, i.e.,  determined by the parameters $k$, $m$ and $N$, is limited by the capability to realize significant number of small width slits, e.g., in micrometer scale, on an appropriate planar surface such as by patterning metallic slits on glass substrate \cite{magana2016exotic}. The beam width and inter-planar distances should be adapted for spatial coherence of the light diffracting through planes \cite{gulbahar2018quantum}. However, the linear modeling of the architecture in {\color{black}Fig.} \ref{Fig5}(a) and the expansion of light beam through propagation allow to realize a feasible architecture in future experimental implementations.

On the other hand, achieving QS experiments allows to adapt certified random number generation protocols for the QPC architecture \cite{aaronson2019, whyte2019,  bouland2019complexity, brakerski2018cryptographic}. Although there are recent high speed, e.g., on the orders of several Gbit/s, random number generation protocols working in a local manner and exploiting the sampling of interference based intensity fluctuations  of laser pulses such as Refs. \citen{rng} and \citen{rng2}, the idea of randomness extraction from QS experiments proposed by Aaronson \cite{aaronson2019} in a way allowing to download from a remote and trusted public source is new. The user interacts with a remote QC and makes it to generate random bits without any trust to the QC itself.  Similar to Aaronson's protocol, it is possible to firstly collect random numbers from a trusted computer. Then, using these numbers, the {\color{black}widths and planar distributions of the slits} are determined to have a random diffraction set-up  by assuming that the mechanical structure of the device can be modified remotely.  Then,  intensity distribution in the photodetector array is measured. Therefore, both the random structure of MPD {\color{black}set-up} and interference of the exponential number of paths result in a very difficult {\color{black}measurement} output intensity to efficiently calculate with classical computers.    However, it is an open issue whether it is possible to utilize the proposed remote QC device based on QPC similar to the Aaronson's protocol which realizes sampling from the n-qubit output of the quantum circuit  and performs Heavy Output Generation {\color{black}(HOG)} tests. On the other hand, QPC does not allow to sample the probability of a single path but the interference of exponentially many number of paths. Therefore, it is a challenge and open issue to adapt the interference sampling operation {\color{black}in QPC} for a similar complexity theoretical proof of randomness generation.  

\subsubsection*{Neuromorphic applications with quantum neuron implementations}

\begin{figure*}[!t]
\centering
\includegraphics[width=6.7in]{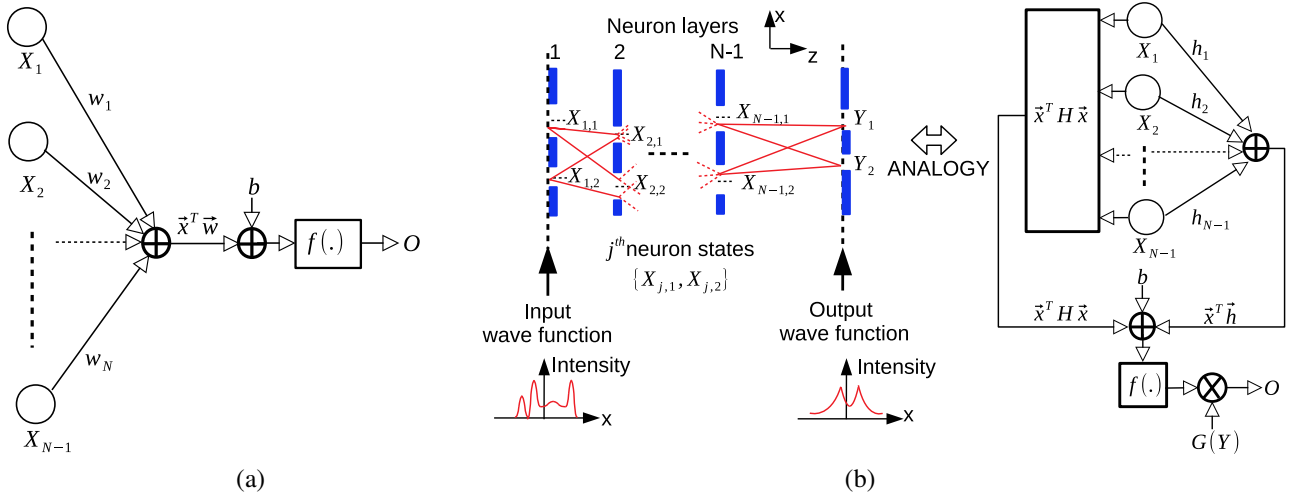} 
\caption{ (a) Classical artificial neuron implementation with the weight and input vectors of $\vec{w}$ and $\vec{x}$, respectively, activation function $f(.)$, bias $b$ and the output $O$. (b) QPC based design of QNNs with quadratic weighting relationship of $\vec{x}^T \, \mathbf{H} \, \vec{x} \, + \, \vec{x}^T \, \vec{h} \, + \,  b$ and nonlinear activation $f(.) \equiv exp(.)$  having quantum superposition and interference combining the inputs.}
\label{Fig6}
\end{figure*} 

In a classical artificial neuron implementation, the output is represented as $O = f( \sum_{i=1} w_i \, x_i  +b)$ where $f(.)$ is the nonlinear activation function, $w_i$ is the $i$th synaptic weight, $x_i$ is the $i$th  input and $b$ is the bias as shown in Fig. \ref{Fig6}(a). QPC based neuron has inherent nonlinearity with the form of $f(\vec{x}^T \, \mathbf{H} \, \vec{x} \, + \, \vec{x}^T \, \vec{h} \, + \,  b)$ between the input slit position vector $\vec{x}$  and output (O) as shown in Fig. \ref{Fig6}(b) {\color{black} based on the MPD formulation for Gaussian sources in (\ref{Eq_new_7}). The} quadratic weighting $\vec{x}^T \, \mathbf{H} \, \vec{x}$, linear weighting $\vec{x}^T \, \vec{h}$  and bias $b$ are fed into the nonlinear function $f(.) \equiv exp(.)$ {\color{black} for the LCT set-up where $\mathbf{H} \equiv \mathbf{\widetilde{H}}_{N-1, n}^{LCT/G}$,  $\vec{h} \equiv \vec{\widetilde{h}}_{N-1,n} \, x_N$, the bias $b \equiv \log(\widetilde{\Upsilon}_{N, n}^G) \, + \, (\widetilde{A}_{N-1,n} \, + \,  \imath \, \widetilde{B}_{N-1,n} ) \, x_N^2$, $\vec{x} \equiv \vec{x}_{N-1,n}$ and $x \equiv x_N$ is the measurement position.}

The example output through the slits $Y_j$  for $j \in [1,2]$ in Fig. \ref{Fig6}(b) depends on interfering quantum superposition of input combinations as follows {\color{black} while assuming path independent forms of the variables for simplicity, i.e., $\mathbf{H} \equiv \mathbf{H}_{N-1}^{LCT/G}$, $\vec{h} \equiv \vec{h}_{N-1} \, x_N$ and $b \equiv \log(\Upsilon_{N}^G) \, + \, (A_{N-1} \, + \,  \imath \, B_{N-1} ) \, x_N^2$}:
\begin{equation}
\label{Eq_new_22} 
O  \equiv \sum_{y \in [Y_1, Y_2]} \, {\color{black}G(y, \beta_{N}, x_N)} \, \sum_{x_1 \in [X_{1,1}, X_{1,2}]} \, \sum_{x_2 \in [X_{2,1}, X_{2,2}]} \hdots \sum_{x_{N-1} \in [X_{N-1,1}, X_{N-1,2}]} f(\vec{x}^T \, \mathbf{H} \, \vec{x} \, + \, \vec{x}^T \, \vec{h} \, + \,  b)
\end{equation}
where {\color{black}$G(y, \beta_{N}, x_N)$}  is the slit mask function depending on the output slit ${\color{black} y \, = \, }Y_i$ for $i \in [1, 2]$ of the $N$th plane {\color{black} and $\beta_{N}$ is the fixed slit width with path-independent assumption for simplicity. The parameters are possible to depend on each path with variable slit masks.} Exponentially large number of synaptic chains (paths) through slit inputs, their quantum interference and simplicity to sample the intensity output {\color{black}$ \vert O \vert^2$} provide significant opportunities to exploit for quantum advantages.

The challenges include designing the quantum neuron based on the slit positions as inputs while changing the weight in a controllable manner. Besides that, extensive simulation studies are required to practically observe the quantum advantages for various problems. The positions of the slits are required to be modified dynamically with special designs. In addition, large scale QNN implementations both in simulations and {\color{black}experiments} are required to observe the performances in various problems, e.g., pattern recognition or machine learning for very large problem sizes.

\subsubsection*{Solution of nonlinear Schr{\"o}dinger equation}

Finite-band solutions of NLSE in (\ref{Eq_new_1}) are expressed with RTF as follows \cite{wahls2015fast, tracy1988nonlinear}:
\begin{equation} 
\label{Eq_new_23}  
q(x,t) = q(x_0, t_0) \, e^{\imath \, k_0 \, x  \, - \, \imath \, \omega_0 \, t} \frac{ {\color{black}\Theta} \big(- \imath \, \mathbf{Y}, \imath \, \frac{\pi}{2}  (\vec{k} \, x \, + \, \vec{\omega} \, t +  \vec{\delta}^-)  \big) }{{\color{black}\Theta} \big(- \imath \, \mathbf{Y}, \imath \, \frac{\pi}{2}  ( \vec{k} \, x \, + \, \vec{\omega} \, t +  \vec{\delta}^+)  \big) }
\end{equation}
where Riemann spectrum is $(\mathbf{Y}, \vec{k},  \vec{\omega}, \vec{\delta}^-, \vec{\delta}^+)$, Riemann period matrix $\mathbf{Y}$  \cite{kamalian2016periodic}  {\color{black}is} calculated using $ \vec {k}$  and $ \vec {\omega} $ together with nonlinear spectrum data  (Appendix to Section 24 in Ref. \citen{osborne2002nonlinear}) and the partial sum of RTF {\color{black} denoted as $\Theta_M$ converging to $\Theta$  for $M \rightarrow \infty$}  is defined as follows:
\begin{align}  
\label{Eq_new_24} 
\begin{split}
   \Theta_M(\mathbf{{\color{black}\Gamma}}, \vec{y}) \equiv 
   \sum_{a_1 = -M}^{M}\hdots \sum_{a_{N-1}= -M}^{M} e^{  - \,\pi \, \vec{a}^T \,  \mathbf{{\color{black}\Gamma}}   \, \vec{a} }   \, e^{2 \, \pi \, \vec{y}^T   \vec{a}}            & 
\end{split}
\end{align}
where $\mathbf{{\color{black}\Gamma}}$  is a  complex matrix, $\vec{y}$ is a complex vector {\color{black}and $\vec{a}^T \equiv [a_1 \,\, a_2 \,\, \hdots a_{N-1}]$}. As shown in Ref. \citen{gulbahar2019quantumpath}, if $ j \in [1, N-1] $ and $ a_j \in S_M \equiv [-M, M]$, we select $ X_ {j, i} \in S_M \Delta x_j$ and also if the {\color{black}slit} widths are kept constant for each plane, then $ A_{\vec{s}}$, $ B_{\vec{s}} $, $ \Upsilon_{\vec{s}} $, $ \mathbf{H_{\vec{s}}} $, $\vec{c}_{\vec{s}} $ and  $ 
\vec{d}_{\vec{s}} $ values become path independent, {\color{black} i.e., $A_{N-1}$, $B_{N-1}$, $\Upsilon_{N}$, $\mathbf{H}_{N-1}$, $\vec{c}_{N-1}$ and $\vec{d}_{N-1}$ while the superscript $(.)^{LCT/G}$ removed for simplicity,} and (\ref{Eq_new_2}) is transformed as {\color{black}follows for  $ x \equiv k  \, T_s $}: 
\begin{align}  
\label{Eq_new_25} 
\begin{split}
\frac{I[k]}{e^{2 \, A_{\color{black}N-1}  \, k^2 T_s^2} \,\vert \Upsilon_{\color{black}N}  \vert^2 \,}  =   {\color{black}\Bigg \vert \Theta_M \Big( \frac{\mathbf{\widehat{\Gamma}} \, + \, \mathbf{\widehat{\Gamma}}^T}{2},  \frac{ x  \,  \mathbf{D}  \,(\vec{c}_{\color{black}N-1} \, + \, \imath \, \vec{d}_{\color{black}N-1})}{2 \, \pi} \Big) \Bigg \vert^2}   & 
\end{split}
\end{align}  
where $  {\color{black}\mathbf{\widehat{\Gamma}}}  \equiv  - \, \mathbf{D} \, \mathbf{H}_{\color{black}N-1} \, \mathbf{D} \, / \, \pi $ and the diagonal matrix $\mathbf{D}$   {\color{black}is} formed of the elements $\lbrace \Delta x_1, \, \Delta x_2, \hdots, \Delta x_{N-1}\rbrace$.    In other words, we can achieve the absolute value of the {\color{black}partial sum of particular RTF} by using the measurement result on the sensor plane. Although the calculation of RTF function is classically quite difficult \cite{gulbahar2019quantumpath, osborne2002nonlinear}, it has important applications in areas including  geometry, arithmetic and number theory \cite{mumford2007tata}, nonlinear spectral theory for ocean and water sciences \cite{osborne2002nonlinear}, cryptography and the solution modeling of NLSEs \cite{wahls2015fast, tracy1988nonlinear}.

{\color{black}Measurements in the QPC system allow to obtain information about}  $ \vert q(x,t)  \vert $. The most important challenge for utilizing QPC in the solutions of NLSE is to determine the set equation parameters $(\mathbf {Y}, \vec{k}, \vec {\omega}, \ \vec {\delta}^{\color{black}\pm})$ which can be implemented with specific QPC design. Theoretical modeling and extensive simulations are required to examine all the practical sample parameters and systems in the literature where NLSE solutions are achieved with RTF based solution.  

Next, numerical simulations are achieved to analyze the effects of {\color{black}FO based} components such as lenses and HG sources on the intensity distribution obtained with QPC. Simulation studies for large scale implementations of QPC for the future applications are open issues.

\section*{Numerical Results}
\label{sec05}

\begin{figure}[!t]
\centering
\includegraphics[height=2.0in]{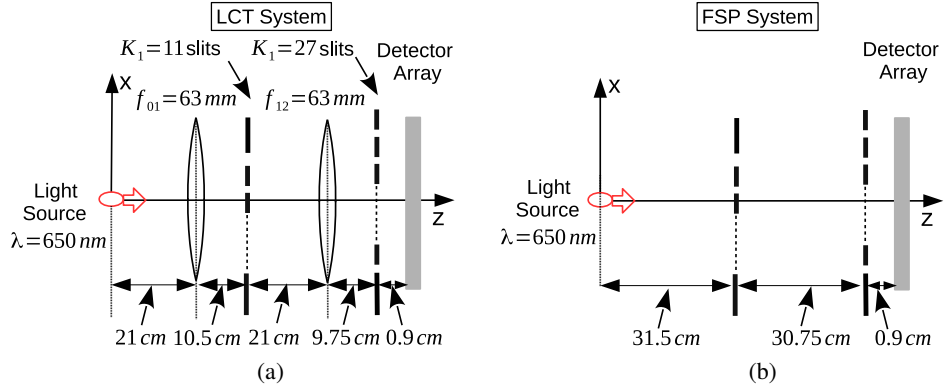} 
\caption{Photonic QPC set-up with Gaussian and HG classical monochromatic light source with $\lambda = 650$ nm, two planes for diffraction with the number of slits $K_1 = 11$ and $K_2 = 27$, respectively,  and specific set-up of (a) LCT and (b) FSP design with the only difference of the existence of thin-lenses of focal length of $63$ mm in the LCT system. }
\label{Fig7}
\end{figure} 
  
\begin{figure}[!t]
\centering
\includegraphics[width=6.55in]{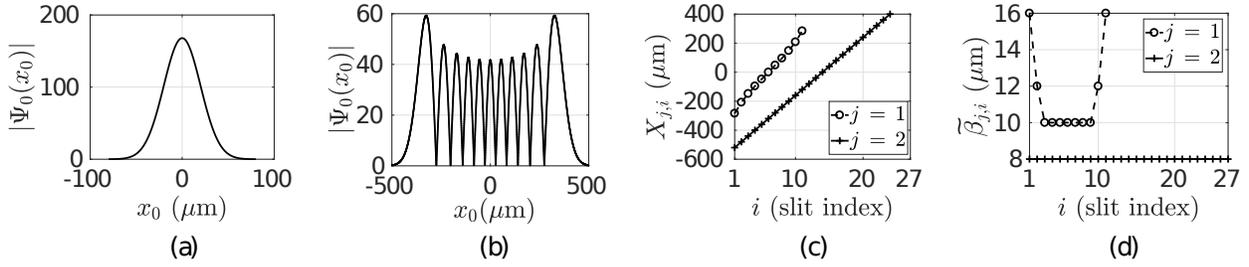} 
\caption{(a) Gaussian and (b) Hermite-Gaussian (order $l = 10$) source waveforms with $\sigma_0 = 20 \, \mu$m and  $W_0 = 200 \, \mu$m, respectively. (c) Slit positions with $K_1 = 11$ and $K_2 = 27$ slits (d) and the widths  $\widetilde{\beta}_{j,i}$ for $j \in [1, 2]$ and $i \in [1, K_j]$.}
\label{Fig8}
\end{figure}  
 
MPD set-up with two diffraction planes and single sensor plane is numerically analyzed for both Gaussian and HG sources with beam width and waist sizes of $\sigma_0 = 20 \, \mu$m and  $W_0 = 200 \, \mu$m, respectively. The set-up is shown in Fig. \ref{Fig7} with $N = 3$ planes. The source waveforms are shown in Figs. \ref{Fig8}(a) and (b), respectively. HG order is set to $l = 10$ with highly oscillatory and negative initial $V_0$ of $1.076$.  The wavelength of the light is chosen in the red spectrum of $\lambda = 650$ nm while the low cost laser sources are commercially available in a wide spread manner. 

Two different set-ups composed of LCT and FSP systems as shown in Figs. \ref{Fig7}(a) and (b), respectively, are compared  where the LCT system includes thin lenses between the diffraction planes while not included in the FSP system. The kernel based on HO in (\ref{Eq_new_33}) giving FRFT as a special case of general LCT  formulation  is not numerically analyzed since LCT based system is a more general version while various combinations including FRFT systems are applicable with the formulation in Tables \ref{Table_Gaussian} and \ref{Table_HG} in the Methods section. Therefore, two different set-ups with the kernels $K_{FS}$ in (\ref{Eq_new_34}) and $K_{LCT}^{a,b,c,d}$ in  (\ref{Eq_new_3}) for the inter-plane propagation are compared for the same design of {\color{black}set-up in terms of the properties of the slits and planes}.  Inter-plane distance vector is given by  $\vec{L}^T \equiv [ 31.5  \, \, \, 30.75 \,\,\, 0.9]$ (cm). The distances of the first plane to the first lens and the second plane to the second lens are denoted with the vector $\vec{L}_a^T \equiv [L_{a,01} \,\,\, L_{a, 12}]$ where both the distances are set to $21$ cm  while the lenses of the focal length $\vec{f}^T \equiv [f_{01} \,\,  f_{12}] = [63 \,\, 63]$ mm focus  the light intensity to  more compact areas on the consecutive planes compared with  FSP. It is assumed that the propagation between the second and third planes includes only FSP without any thin lens to simplify the set-up. $K_1 = 11$ and $K_2 = 27$ slits are used on the first and second planes, respectively. The slit positions and widths on the first plane,  as shown in Figs. \ref{Fig8}(c) and (d), respectively, are adapted to the maximum intensity locations of HG source propagation on the first plane in LCT system while the ones on the second plane are chosen uniformly with the separation of $40 \, \mu$m and the width of $\widetilde{\beta}_{2, j}  \equiv \beta_2 = 8 \, \mu$m. $K_{LCT}^{a_{01}, \, b_{01}, \, c_{01}, \,d_{01}}$  and $K_{LCT}^{a_{12}, \, b_{12}, \, c_{12}, \, d_{12}}$  are calculated by using (\ref{Eq_new_5}). 

FSP has less control over the propagation of light compared with LCT based {\color{black}FO}. FSP spreads the light without any tuning to the slit positions by reducing the probability of the {\color{black}photon} to reach to the consecutive planes after diffraction. Therefore, in numerical analysis, LCT is shown to improve the probability of {\color{black}photon} detection on the sensor plane ($P_E$) and also the negative volume of Wigner function  compared with FSP. The vectors of $P_E(j)$ and $V_j$  composed of the values on the first, second and third (sensor) planes for Gaussian sources are denoted with $\vec{P}_{E}^{G, FSP}$ and $\vec{V}^{G, FSP}$, respectively, for FSP while with $\vec{P}_{E}^{G, LCT}$ and $\vec{V}^{G, LCT}$ for LCT. The cases with HG sources are denoted with the superscript of $HG$. It is an open issue to adapt LCT parameters with respect to any given set-up including inter-plane distances, slit locations and widths in a way to maximize the interference and the probability of the {\color{black}photon} reaching to the sensor plane.
 
\begin{figure*}[!t]
\centering
\includegraphics[width=6.9in]{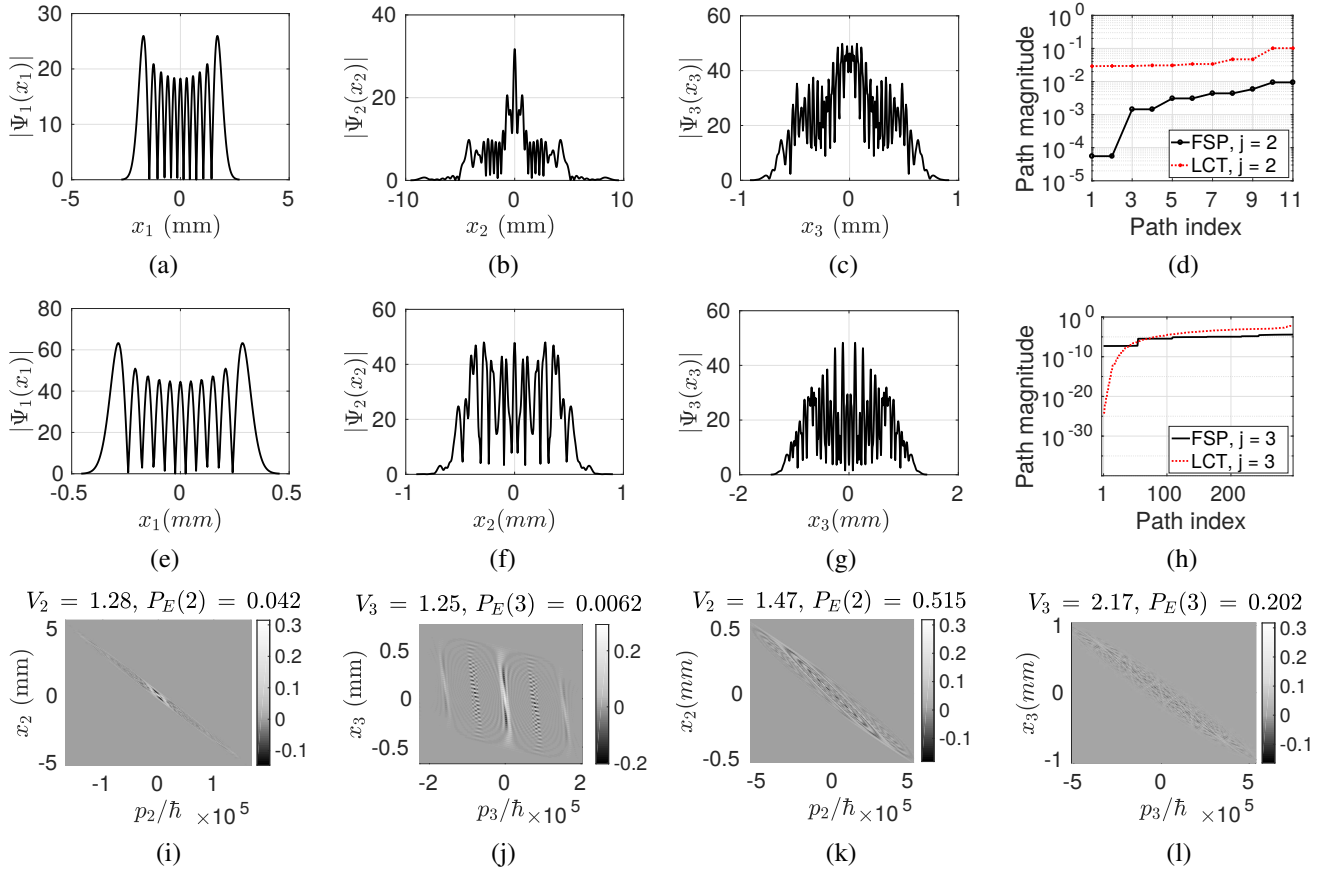} 
\caption{HG source with the order of $l =10$ and $W_0 = 200 \, \mu$m is utilized where the resulting spatial  domain waveforms  on the planes with the indices (a) $j = 1$, (b) $j = 2$ and (c) $j = 3$ for FSP, and (e) $j = 1$, (f) $j = 2$ and (g) $j = 3$ for LCT.  $P_E(n, j)$ for (d) $j =2$ and (h) $j = 3$. Scaled Wigner distribution $\hbar \times W_j(x_j, p_j)$ for FSP on the planes with (i) $j = 2$  and (j) $j = 3$, and for LCT with (k) $j = 2$  and (l) $j = 3$.}
\label{Fig9}
\end{figure*}  

\begin{figure*}[!t]
\centering
\includegraphics[width=6.8in]{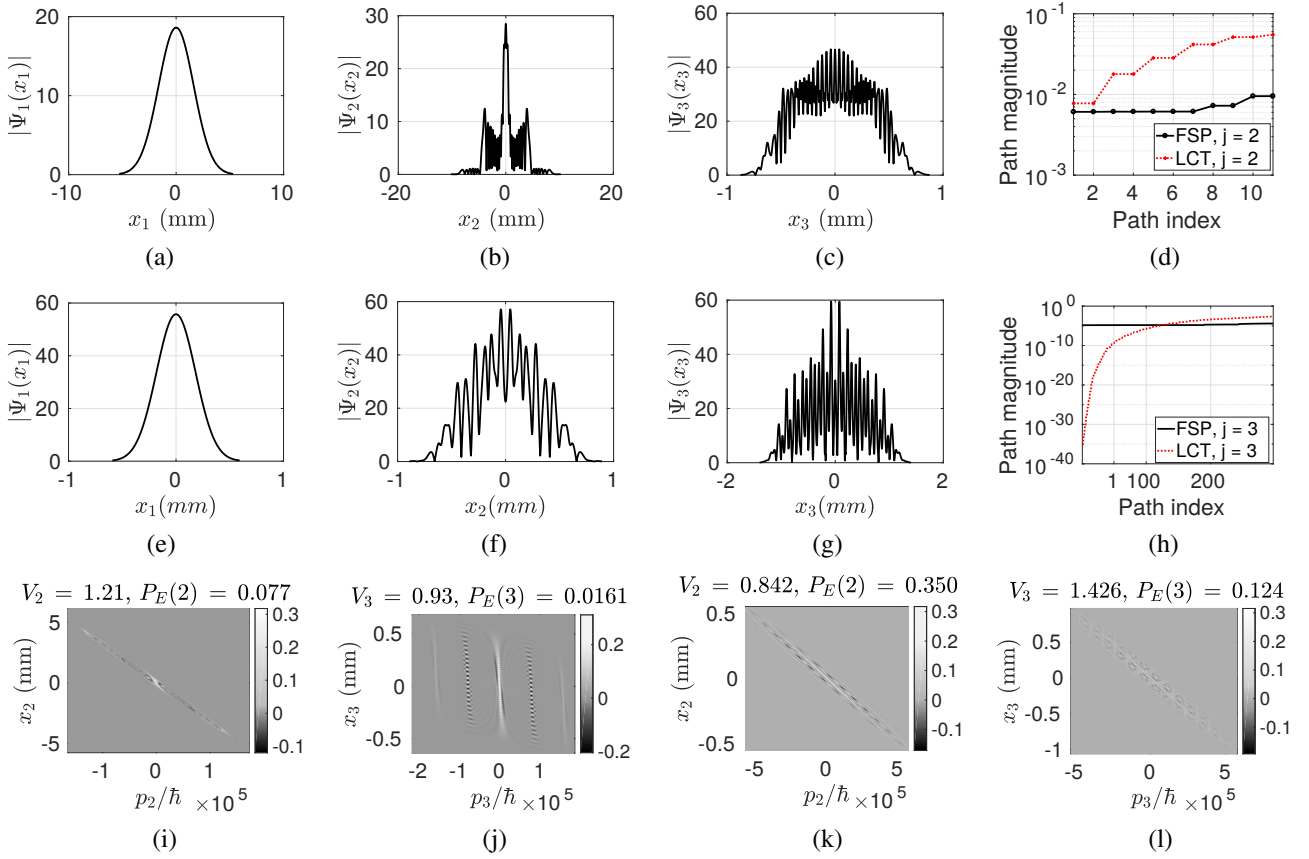} 
\caption{Gaussian source with   $\sigma_0 = 20  \, \mu$m is utilized where the resulting spatial  domain waveforms  on the planes with the indices (a) $j = 1$, (b) $j = 2$ and (c) $j = 3$ for FSP, and (e) $j = 1$, (f) $j = 2$ and (g) $j = 3$ for LCT.  $P_E(n, j)$ for (d) $j =2$ and (h) $j = 3$. Scaled Wigner distribution $\hbar \times W_j(x_j, p_j)$ for FSP on the planes with (i) $j = 2$  and (j) $j = 3$, and for LCT with (k) $j = 2$  and (l) $j = 3$.  }
\label{Fig10}
\end{figure*} 

\subsection*{Hermite-Gaussian Sources}
\label{sec0501}

The waveforms on the three planes in spatial domain  are shown in Figs. \ref{Fig9}(a), (b), (c) and  \ref{Fig9}(e), (f), (g) for FSP and LCT cases, respectively. It is observed that LCT focuses the light better on the slit locations while FSP reduces $P_E$ significantly.  $\vec{P}_{E}^{HG, LCT} = [1 \,\,  0.515 \,\,  0.202]$ is much more improved compared with $\vec{P}_{E}^{HG, FSP} = [1 \,\,  0.042 \,\,  0.0062]$. The magnitudes of the interfering paths  are shown in  Figs. \ref{Fig9}(d) and (h), for $j =2$ and  $j = 3$, respectively, while Wigner distributions on the second and third planes scaled with  $\hbar$ are shown in Figs. \ref{Fig9}(i) and (j) for FSP and, (k) and (l) for LCT. It is observed that LCT provides significantly larger path magnitudes while $\vec{V}^{HG, LCT} = [1.076 \,\,  1.47 \,\,  2.17]$ is also improved compared with $\vec{V}^{HG, FSP} = [1.076 \,\,  1.28\,\,  1.25]$. Observe that HG source has already negative Wigner volume of $1.076$ which is much further improved by LCT set-up compared with  FSP.

\subsection*{Gaussian Sources}
\label{sec0502}
  
The waveforms in spatial domain for Gaussian sources are shown in Figs. \ref{Fig10}(a), (b), (c)   and \ref{Fig10}(e), (f), (g)    for FSP and LCT cases, respectively.  $\vec{P}_{E}^{G, LCT} = [1 \,\,  0.35 \,\,  0.124]$ and  $\vec{P}_{E}^{G, FSP} = [1 \,\,  0.077 \,\,  0.0161]$ values are obtained where the magnitudes of the interfering paths  are shown in  Figs. \ref{Fig10}(d) and (h), for $j =2$ and  $j = 3$, respectively.  Similar to the HG sources, LCT improves diffraction probabilities significantly compared with FSP. Wigner distributions scaled with  $\hbar$ are shown in Figs. \ref{Fig10}(i), (j), (k) and (l) having different characteristics compared with HG case in  Figs. \ref{Fig9}(i), (j), (k) and (l). It is similarly observed that LCT provides significantly larger path magnitudes. $\vec{V}^{G, LCT} = [0 \,\,  0.842 \,\,  1.426]$ and $\vec{V}^{G, FSP} = [0 \,\,  1.21\,\,  0.93]$ are obtained with increasing interference complexity through diffraction on consecutive planes in LCT case while starting with purely classical Gaussian source of zero negative Wigner volume.  $V_2$ of $0.842$ for LCT is smaller than $1.21$ for FSP on the second plane. This is due to the both the specific set-up parameters and more diverse distribution of the path magnitudes in LCT after diffraction from the first plane as shown in Fig. \ref{Fig10}(d). It becomes more difficult on the third plane to correlate the distribution of the path magnitudes shown in Fig. \ref{Fig10}(h) with $V_N = V_3$ shown in Figs. \ref{Fig10}(j) and (l). In other words, complexity behaves differently compared with the transmission probability while requiring simultaneous maximization depending on specific set-up as an open issue as discussed in the Results section while presenting Wigner distribution analysis.   

\section*{Open Issues and Discussion}
\label{sec06}

There are some open issues to best exploit {\color{black} photonic QPC method based on MPD  and FO}. Mathematical formulation correlating specific set-up parameters to  path magnitude distribution and  negative volume of Wigner function is an open issue. Iterative formulation of the vectors $\vec{\widetilde{h}}_{N-1,n}$, $\vec{\widetilde{\gamma}}_{N-1,n}$, $\vec{\widetilde{\eta}}_{N-1,n}$ and the matrices $\widetilde{\mathbf{H}}_{N-1, n}^{HO/G}$, $\widetilde{\mathbf{H}}_{N-1, n}^{LCT/G}$ and $\widetilde{\mathbf{H}}_{N-1, n}^{LCT/HG}$  in (\ref{Eq_new_7}), (\ref{Eq_new_11}) and (\ref{Eq_new_13}) are complicated as shown in the Methods section with complicated parameters and iterations in Tables \ref{Table_Gaussian} and \ref{Table_HG}. Therefore, adapting the physical set-up parameters to the desired form of partial sum of {\color{black}RTF} for the target number theoretical problem, and  characterizing the path distributions and the negative volume of Wigner function explicitly are important open issues.

{\color{black} In the proposed formulation, the qubits are obtained through the tensor product structure of projections at different time instants on the contrary with the spatial encoding and entanglement of multiple photons. However, the histories of photon trajectories are not formulated for realizing conventional quantum gate implementations. Implementations of the quantum circuit gates are required to obtain universal QC architectures. Therefore, proposed QC formulation is limited to utilization of the interference of exponentially increasing number of Feynman paths based on the superposition and coherence properties of light source. Implementations of quantum circuits and fundamental search algorithms such as Grover search are future works to clarify the potential future scope of QPC based computing architectures in terms of universal QC capabilities. On the other hand, {\color{black}FO} based QPC implementation  has two main advantages resembling Boson sampling advantages in a different context \cite{aaronson2011computational}: (a) not utilizing multiple photons as qubits getting rid of the coupling disadvantages while exploiting single photon trajectories and (b) utilizing the \textit{free entanglement}, for the first time, of the \textit{classical light} obtained through freely available temporal correlations among the projections at different time instants. Boson sampling compared with QPC utilizes still multiple indistinguishable photons (but not as qubits) while requires single photon generation and detection to exploit the free entanglement among indistinguishable photons through multi-mode interferometer with the regarding boson statistics.}

Realizing perfectly Gaussian slits compared with the conventional rectangular apertures is an important open issue for matching the experimental results with the proposed theoretical model. However, any slit structure can be represented as a composition of Gaussian slits by using the method defined in Ref. \citen{wen1988diffraction} and applied successfully in optical diffraction theory and experiments \cite{ding1999approximate, lu2014experimental}. The one dimensional slit {\color{black}mask} function ${\color{black}\widehat{G}}(x)$ is represented as follows: 
\begin{equation}
\label{Eq_new_26}
{\color{black}\widehat{G}}(x) \approx \sum_{i =1}^{K} a_i \, \exp( - \, x^2 \, / \, 2 \, \beta_i^2) 
\end{equation} 
where $a_i$ and $\beta_i$ are found with optimization based on the experimental measurement results while increasing $K$ provides more accurate results. If the perfect Gaussian slits are replaced with the superposition in (\ref{Eq_new_26}),   then the summations in  (\ref{Eq_new_7}), (\ref{Eq_new_11}) and (\ref{Eq_new_13})  should be made for each $\beta_i$ of the single slit. The functional form with partial sum of {\color{black}RTF} should be calculated and summed for each combination of $\beta_i$ through all the slits. Therefore, non-Gaussian slits {\color{black}can possibly} realize the solutions of much harder computational complexity problems {\color{black}as an open issue}. 

There are some factors {\color{black}effecting the degree of} compatibility between the theory and practice. These include imperfection in optical set-up, e.g., finite size lens effects, planar thickness, characterization of slit functions, sources and detector efficiency. The theoretical model should be extended including all the set-up parameters having diverging effects on the final intensity distribution. Similarly, the effects of exotic paths, i.e., trajectories between the slits on the same plane, should be included in the mathematical model as thoroughly discussed in Refs. \citen{sawant2014nonclassical, da2016gouy, gulbahar2019quantumpath}. All these considerations potentially lead to unavoidable errors requiring quantum error correction studies adapted to QPC architectures \cite{kitaev1997quantum}. {\color{black}Another open issue is related to the utilization of the measurements on all the sensor planes for computational purposes not only the final sensor plane since they include diffraction through previous planes. Theoretical models are required to exploit the sensor measurement results.}

\section*{Methods}

\subsection*{Quantum Fourier Optics}
\label{sec02}

\begin{figure}[!t]
\centering
\includegraphics[width=1.9in]{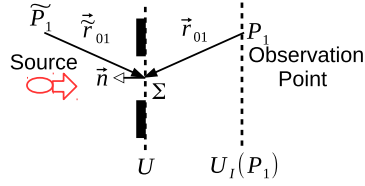}  
\caption{Formulation set-up for first type of solution of Rayleigh-Sommerfeld diffraction through a slit     $\Sigma$ \cite{goodman2005introduction}.}
\label{Fig11}
\end{figure} 
 
In scalar diffraction theory, the first Rayleigh-Sommerfeld formula of the Huygens-Fresnel principle for the propagation of light on planar surfaces is described as follows by using the Green's theorem \cite{goodman2005introduction}:
\begin{equation}
\label{Eq_new_27}
 U_I(P_{1}) = \frac{-1}{4 \, \pi} \iint\limits_{\Sigma} U \, \frac{\delta G_{-}}{\delta n} \, ds
\end{equation} 
where $U_I(P_{1})$ is the wave amplitude at the point $P_{1}$, $U$ is the distribution on the planar screen where diffraction occurs, $\Sigma$ denotes the integration over the slit including its multiplicative effects on the wave amplitude, $G_{-} \equiv exp(\imath \, k \, r_{01}) \, / \, r_{01} - exp(\imath \, k \, \widetilde{r}_{01}) \, / \, \widetilde{r}_{01}$ is the Green's function vanishing on the diffraction surface for the first type of solution of Rayleigh-Sommerfeld formula, $r_{01} \equiv \vert \vec{r}_0 - \vec{r}_1 \vert$  and $ k \equiv 2 \, \pi \,/ \, \lambda $ for the monochromatic light source of wavelength $\lambda$ as shown in Fig. \ref{Fig11} \cite{goodman2005introduction}. Assuming that $r_{01} \gg \lambda$, the following approximation holds in rectangular coordinates:
\begin{equation}
\label{Eq_new_28}
 U_I(P_{1}) \approx  \iint\limits_{\Sigma} U(x_{0}, y_{0}) \, K_{FS}(\vec{r}_1, \vec{r}_0) \,dx_{0} \, dy_{0}
 \end{equation}
where the kernel $K_{FS}(\vec{r}_1, \vec{r}_0)$ for FSP is defined as follows:
\begin{eqnarray}
\label{Eq_new_29}
K_{FS}(\vec{r}_1, \vec{r}_0)  & \equiv & \frac{1}{ \imath \, \lambda}  \, \frac{e^{\imath \, k \, r_{01}}}{r_{01}} \, \cos(\vec{n}, \vec{r}_{01})  \\
\label{Eq_new_30}
 & = & \frac{z}{ \imath \, \lambda}  \, \frac{e^{\imath \, k \, r_{01}}}{r_{01}^2}  \\
 \label{Eq_new_31}
 & \approx & \frac{ e^{\imath \, k \,z}}{\imath \, \lambda \, z}   e^{\frac{\imath \, k }{2 \, z} \,\big( (x_{1} - x_{0})^2 + (y_0 - y_{0})^2 \big)  }
 \end{eqnarray}
 where $r_{01} = \sqrt{z^2 \, + \, (x_{1} - x_{0})^2 \, + \, (y_0 - y_{0})^2}$. The kernel for Fresnel diffraction integral is obtained by further approximation of $r_{01}$ in the near-field for large $z$ resulting in (\ref{Eq_new_31}). This expression is the convolution integral conventionally used in phase-space optics for FSP \cite{ozaktas2001fractional}. 
 
Recently, scalar diffraction theory and Fresnel diffraction integral are discussed in Ref. \citen{santos2018huygens} to be validly representing the evolution of light wave function  modeled with the Hamiltonian of the quantized electromagnetic field  $H =  (\hat{p}^2 + \omega^2 \,\hat{q}^2) \, / \, 2$ as the Feynman's path integral (FPI) solution of the quantum mechanical   HO  \cite{feynman2010quantum}. Fresnel diffraction nature of the propagation is verified with experimental photon counting studies for single photons. The wave function amplitude of light field in one dimension  on a plane $\Psi(x_0)$ is modeled to propagate  into the amplitude $\Psi(x_1)$ on another plane (Eq. 16 in Ref. \citen{santos2018huygens} transformed into a simpler form) with the following formulation:
\begin{equation}
\label{Eq_new_32}
\Psi(x_{1}) \approx  \int_{-\infty}^{\infty} \Psi(x_{0}) \, K_{HO}(x_{1}, x_{0}) \,dx_{0}  
\end{equation}
where the kernel based on HO is the following: 
\begin{align}
\begin{split}
\label{Eq_new_33} 
K_{HO}(x_{1}, x_{0}) \equiv  \sqrt{ \frac{m_{\lambda}}{ 2 \, \pi \, \imath \, \hbar \,t_{01}\, \sin(\omega \, t)}}   \,\exp\bigg(\frac{\imath \, m_{\lambda} \, \big( x_{1}^2 \, \, \cos(\omega \, t) \, - \, 2 \, x_{1} \, x_{0} \, + \, x_{0}^2  \, \cos(\omega \, t) \big) }{2 \, \hbar \, t_{01} \,  \sin(\omega \, t)} \bigg) & 
\end{split}
\end{align} 
where  $c$ is the velocity of light, $\omega \equiv 2 \, \pi \, c \, / \, \lambda$,  $ \omega \, t \neq n \, \pi$ for $n \in \mathbb{Z}$, $t_{01}$ is the propagation duration between the planes and $m_{\lambda} \equiv \hbar \, k \, / \, c$ is the defined equivalent mass of photon propagation. In addition, the approximated FSP kernel in (\ref{Eq_new_31}) is simply converted to the following in 1D system:  
\begin{equation}
\label{Eq_new_34}
K_{FS}(x_1, x_0) \approx e^{j \, k \,z} \sqrt{\frac{m_{\lambda} }{2 \, \pi \, \imath\, \hbar \, t_{01}  }  } e^{\frac{\imath \, m_{\lambda}  }{2 \, \hbar \,t_{01} } \,  (x_{1} - x_{0})^2}
\end{equation}  
The kernel for massive particles with the mass $m$ such as an electron is expressed as follows \cite{feynman2010quantum, gulbahar2019quantumpath}:
\begin{equation}
\label{Eq_new_35}
K_{m, FS}(x_1, x_0) \approx   \sqrt{\frac{m }{2 \, \pi \, \imath\, \hbar \, t_{01}  }  } e^{\frac{\imath \, m }{2 \, \hbar \,t_{01} } \,  (x_{1} - x_{0})^2}
\end{equation}
In other words, the formulation based on phase-space optics for photon and electron propagation wave amplitudes have the similar form in (\ref{Eq_new_34}) and (\ref{Eq_new_35}) except an overall phase factor. The form in (\ref{Eq_new_34}) is utilized in Ref. \citen{gulbahar2019quantumspatial} for defining QSM while targeting only classical communications.
  
On the other hand, both the kernels $K_{HO}(x_{1}, x_{0})$ and $K_{FS}(x_1, x_0)$ are special cases of LCTs  defined for quadratic-phase optics \cite{ozaktas2001fractional}. As a class of linear integral transforms, they include as special cases the Fresnel {\color{black}transform and  FRFT}, simple scaling, chirp multiplication and some other operations. Spatial distribution of light in phase-space optics for the class denoted by quadratic-phase systems is mathematically equivalent to  LCTs (Chapters 3 and 8 in Ref. \citen{ozaktas2001fractional}). These optical systems include arbitrary combinations of the sections of free space in the Fresnel approximation, thin lenses and   sections of quadratic graded-index media. In Ref. \citen{santos2018huygens}, FRFT nature of the kernel $K_{HO}(x_{1}, x_{0})$ is shown both theoretically and experimentally while emphasizing the applicability of all the properties of Fourier analysis to quantum optics. In this article,   propagation of the wave function is extended to the general case of LCTs providing  flexibility to utilize arbitrary optical set-ups by enlarging the functional structures and number theoretical problems exploited in QPC. Furthermore, a better control is obtained for the energy flow of the light through the slits. 

The kernel matrices for $K_{HO}(x_{1}, x_{0})$ and $K_{FS}(x_1, x_0)$ are given as follows:
\begin{eqnarray}
\label{Eq_new_36}
\mathbf{M}_{HO} & = & \begin{bmatrix}
\cos(\omega \, t)& \frac{ 2 \, \pi \, \hbar \,t_{01}\, \sin(\omega \, t)}{m_{\lambda}}\\
-\frac{ m_{\lambda} \, \sin(\omega \, t)}{2 \, \pi \,  \hbar \,t_{01}} &\cos(\omega \, t)\\
\end{bmatrix} \\
\label{Eq_new_37}
\mathbf{M}_{FS}  & = & \begin{bmatrix}
1& \frac{ 2 \, \pi \, \hbar \,t_{01}}{m_{\lambda}} \\
0 &1\\
\end{bmatrix} 
\end{eqnarray}
$\mathbf{M}_{HO}$ has the same form with the propagation of light in quadratic graded-index media of having the refractive index distribution of $n^2(x) =  n_0^2(1 - (x \, / \, \chi)^2)$ where   $n_0$ and $\chi$ are the medium parameters. The parameter matrix of the propagation through the quadratic graded-index medium of length $d_{gri}$ is given by the following (Section 8.3.3 in Ref. \citen{ozaktas2001fractional}):
\begin{equation}
\label{Eq_new_38}
\mathbf{M}_{gri}   =  \begin{bmatrix}
\cos(\alpha)&   \lambda_{\chi} \, \sin(\alpha) \\
- \sin(\alpha) \, / \, \lambda_{\chi} & \cos(\alpha)\\
\end{bmatrix} 
\end{equation}
where  $\alpha = d_{gri} \, / \, \chi$. There is a FRFT relation between scaled versions of the input  $\hat{f}(x)$ and output $\hat{g}(x)$ with FRFT order $\alpha$ as $\hat{g}(x) = e^{- \imath \, d_{gri} \, / \, (2 \,  \chi)} \lambda_{\chi}^{ -1 \,/ 4 } f_a( x \, / \, \sqrt{\lambda_{\chi}})$ where $f(x) \equiv \lambda_{\chi}^{1 \,/ 4} \, \hat{f}(x \, \sqrt{\lambda_{\chi}})$ and $f_a(x)$ denotes the $a$th order FRFT of $f(x)$. FRFT operation of order $\alpha$ is represented with the parameter matrix of $a = d = \cos(\alpha)$ and $b = \sin(\alpha)$. As a result, $\mathbf{M}_{HO}$ represents a FRFT relation between the  input and output scaled with the parameter $ \sqrt{\lambda_{\chi}}$ where the parameters are $\alpha = \omega \, t$ and $\lambda_{\chi} \equiv   2 \, \pi \, \hbar \,t_{01} \, / \, m_{\lambda}$ while as a special case of LCTs. 

\subsection*{Matrix formulation for HO/LCT System with Gaussian sources}
\label{appa}

The following formulation is   valid for both HO and  LCT based design with Gaussian sources {\color{black}where} the corresponding {\color{black}iteration} parameters {\color{black}are}  defined in Table \ref{Table_Gaussian}. The elements in the vector $\vec{\widetilde{h}}_{N-1,n} = \vec{\widetilde{c}}_{N-1, n} \, + \, \imath \, \vec{\widetilde{d}}_{N-1, n} $ are defined as  follows:
\begin{equation}
\label{Eq_new_39}
 \begin{bmatrix}
\vec{\widetilde{c}}_{N-1, n}^T \\  \vec{\widetilde{d}}_{N-1, n}^T
\end{bmatrix} =  \begin{bmatrix}
\vec{\widetilde{v}}_{0, N-1, n} & \vec{\widetilde{v}}_{1, N-1, n}  & \hdots  &   \vec{\widetilde{v}}_{N-2, N-1, n}
\end{bmatrix}
 \end{equation} where $\vec{\widetilde{v}}_{k, j, n}$ for $k \in [0, j-1]$ is given as follows:
\begin{equation}
\label{Eq_new_40}
\vec{\widetilde{v}}_{k, j, n} \, \equiv \, \bigg(  \prod_{i=1}^{j-1-k}  \begin{bmatrix}
\widetilde{p}_{4, \,j+1-i, \, n}       & \widetilde{p}_{5, \, j+1-i,\, n}  \\ -\widetilde{p}_{5, \, j + 1 - i,\, n}       & \widetilde{p}_{4, \,j+1-i,\, n}   
\end{bmatrix}   \bigg) \begin{bmatrix}
\widetilde{\zeta}_{k+1,c, n}   \\  \widetilde{\zeta}_{k+1,d, n} 
\end{bmatrix}
 \end{equation} 
{\color{black}Here, the} matrix multiplication   $\prod_{i=1}^{k} \mathbf{U}_i$ denotes $\mathbf{U}_1 \, \mathbf{U}_2 \hdots \mathbf{U}_k$ for any matrix $\mathbf{U}_i$ for $i \in [1,k]$ and $\widetilde{p}_{4,j, n}$, $\widetilde{p}_{5,j, n}$, $\widetilde{\zeta}_{j, c, n}$  and $\widetilde{\zeta}_{j, d, n}$ for $j \in [1, N-1]$ are defined in Table \ref{Table_Gaussian}.   Assume that $\mbox{diag}\lbrace \vec{y}_1, \hdots,  \vec{y}_K \rbrace$ and $\mbox{diag}\lbrace \mathbf{y}_1, \hdots,  \mathbf{y}_K \rbrace$ define the operators creating  block diagonal matrices by putting the vectors $\vec{y}_j$ and the matrices $\mathbf{y}_j$ for $j \in [1, K]$, respectively,  (all the vectors or the matrices having the same dimensions) to the main diagonal and making zero the remaining elements. The matrix $\mathbf{\widetilde{H}}_{N-1, n}^{HO/G}$ is more simplified as follows compared with the more complicated form achieved for electron based FSP in Ref. \citen {gulbahar2019quantumpath}:
 \begin{align} 
 \begin{split}
\label{Eq_new_41}
\mathbf{\widetilde{H}}_{N-1, n}^{HO/G}  \, = \,\mathbf{\widetilde{D}}_{a, N-1, n}^{HO/G}      
\, + \,  \begin{bmatrix}
\widetilde{\mathbf{V}}_{N-1,n}^T \,  \mathbf{\widetilde{D}}_{b, N-1, n}^{HO/G} \, \widetilde{\mathbf{V}}_{N-1,n} &   \vec{0}_{N-2}\\
\vec{0}_{N-2}^T  & 0  \\
\end{bmatrix}     
\, + \,  \begin{bmatrix}
\vec{0}_{N-2}^T  & 0  \\
 \mathbf{\widetilde{D}}_{c, N-1, n}^{HO/G} \, \widetilde{\mathbf{V}}_{N-1,n} &   \vec{0}_{N-2}\\
\end{bmatrix}     & \\
\end{split}
\end{align} 
where the diagonal matrices are defined as follows: 
\begin{eqnarray}
\label{Eq_new_42}
\mathbf{\widetilde{D}}_{a, N-1, n}^{HO/G} = \mbox{diag}\lbrace \widetilde{p}_{1, 1, n}, \, \widetilde{p}_{1, 2, n}, \, \hdots,\, \widetilde{p}_{1, N-1, n} \rbrace && \\
\label{Eq_new_43}
\mathbf{\widetilde{D}}_{b, N-1, n}^{HO/G} = \mbox{diag}\lbrace \widetilde{\mathbf{K}}_{2, b, n},   \, \widetilde{\mathbf{K}}_{3, b, n}, \, \hdots,\, \widetilde{\mathbf{K}}_{N-1, b, n} \rbrace && \\
\label{Eq_new_44}
\mathbf{\widetilde{D}}_{c, N-1, n}^{HO/G} = \mbox{diag}\lbrace \widetilde{\vec{k}}^T_{2, c, n},   \, \widetilde{\vec{k}}^T_{3, c, n}, \, \hdots,\, \widetilde{\vec{k}}^T_{N-1, c, n}  \rbrace &&  
\end{eqnarray}
$2 \times 2$ block  $\widetilde{\mathbf{K}}_{ j,  b, n}$ and  $1 \times 2$  vector $\widetilde{\vec{k}}_{ j,   c, n}^T$  for $j \in [2, N-1]$ are defined as follows:
\begin{eqnarray}
\label{Eq_new_45}
\widetilde{\mathbf{K}}_{ j, b, n} \, & =  & \, \frac{\widetilde{\beta}_{j,n}^2 \, \widetilde{p}_{3,j,n}}{2} \, \begin{bmatrix}
 1 &   \imath \\
\imath  &  -1 \\
\end{bmatrix}    \\
\label{Eq_new_46}
\widetilde{\vec{k}}_{ j, c, n}^T \,&  = &\, \widetilde{p}_{3,j,n} \, \begin{bmatrix}
1 &   \imath \\
\end{bmatrix} 
\end{eqnarray} 
$\widetilde{\mathbf{V}}_{N-1, n}$ is a lower triangular block matrix defined as follows:
\begin{equation} 
\label{Eq_new_47}
 \begin{bmatrix}
 \vec{\widetilde{v}}_{0,1, n} & \vec{0}_2 & \vec{0}_2 & \hdots & \vec{0}_2 \\
  \vec{\widetilde{v}}_{0,2, n} &  \vec{\widetilde{v}}_{1,2, n} & \vec{0}_2 & \hdots &   \vec{0}_2 \\
  \vec{\widetilde{v}}_{0,3, n} &  \vec{\widetilde{v}}_{1,3, n} & \vec{\widetilde{v}}_{2,3, n} & \hdots   & \vec{0}_2 \\
\vdots &  \vdots & \vdots &  \ddots & \vec{0}_2  \\
 \vec{\widetilde{v}}_{0,N-2, n} &  \vec{\widetilde{v}}_{1,N-2, n}&  \vec{\widetilde{v}}_{2,N-2, n} &  \hdots &   \vec{\widetilde{v}}_{N-3,N-2} \\
\end{bmatrix}
\end{equation}
Expanding  $\mathbf{\widetilde{H}}_{N-1, n}^{HO/G}$ in terms of real and imaginary parts is achieved by finding the real and imaginary parts of $\widetilde{p}_{1, j, n}$ for $j \in [1, N-1]$ and $\widetilde{p}_{3, j, n}$ for $j \in [2, N-1]$, and $\widetilde{\mathbf{K}}_{ j, b, n}$ and $\widetilde{\vec{k}}_{ j, c, n}$ for $j \in [2, N-1]$ since $\widetilde{\mathbf{V}}_{N-1,n}$ is a real matrix. This is easily achieved by using the explicit forms of $\widetilde{p}_{1, j, n}$ and $\widetilde{p}_{3, j, n}$ in Table \ref{Table_Gaussian}. Some variables and constants used  in Table  {\color{black}\ref{Table_Gaussian} (not defined in the table)}  are the following: $\imath \equiv \sqrt{-1}$, $x_{j,n}$ denotes $X_{j, s_{n, j}}$,  $\lambda_0 \, =  \, \hbar  \, t_{0, 1}$,      $\alpha_j \,= \,\omega \, t_{j, j+1} $ for $j \in [0, N-1]$,      $\widehat{m}_{j} = m_{\lambda} \, / \, \sin( \omega \, t_{j, j+1})$ and $ \widetilde{\lambda}_{j,n}^{*}$ as the conjugate of $\widetilde{\lambda}_{j,n}$ for real values of $(a_{j, j+1}, b_{j, j+1}, c_{j, j+1}, d_{j, j+1})$.

Polynomials for the example in (\ref{Eq_example1}-\ref{Eq_example4}) are presented in Table \ref{Table_Example_Gaussian} for the simple case of $N =3$ and Gaussian source.  It is possible by using the explicit modeling to make various gedanken experiments and to perform complexity theoretical calculations.

 \setlength\tabcolsep{2 pt}    
\renewcommand{\arraystretch}{2.1}
\begin{table*}[!t]
\caption{Iteration parameters for FPI modeling of MPD with the kernels $K_{HO}$ and $K_{LCT}^{(a,b,c,d)}$ for Gaussian sources}
\vspace{-0.2in}
\begin{center}
\footnotesize
%\begin{tabularx}{0.98\linewidth}{|c|m{3.2cm}|c|} 
\begin{tabular}{|m{2.0cm}|m{7.8cm}|m{6.2cm} |}
\hline 
  & Formula for $K_{HO}$ based MPD ($\omega \, t_{j, j+1} \, / \, \pi \not\in  \mathbb{Z}$) &  Formula for $K_{LCT}^{(a,b,c,d)}$ based MPD ($b_{j, j+1} \neq 0$)  \\
\hline 
   \multicolumn{3}{ |c|}{$ \Psi_0(x_0) = \mbox{exp}\big(- \, x_0^2 \, / \, (2 \, \sigma_0^2) \big) $ $/ $ $ \sqrt{\sigma_0 \,\sqrt{\pi }} $ $\&$  $ \Psi_1(x_1) =  \chi_0 \,  \mbox{exp}\big(A_0 \,x_1^2  \, + \, \imath \, B_0 x_1^2 \big)$  }    \\ 
\cline{1-3} 
 $A_0$ &     $ - \,\widehat{m}_{0}^2 \,\sigma_0^2   \, / \, \big(    2 \, \cos^2(\alpha_0) \,\widehat{m}_{0}^2\, \sigma_0^4 \, + \, 2 \,\lambda_0^2   \big) $   &   $- 2 \, \pi^2 \, \sigma_0^2   \, / \, \big( 4  \,\pi ^2  \, a_{01}^2  \,\sigma_0^4 \,+ \, b_{01}^2 \big) $  \\
\cline{1-3} 
 $B_0$ &  $ \cos(\alpha_0) \,\widehat{m}_{0} \, \left(\lambda_0^2  \, -  \,\sin^2(\alpha_0) \, \widehat{m}_{0}^2 \sigma_0^4 \right) \, / \, \big( 2 \, \lambda_0 \,  (\cos^2(\alpha_0) \, \widehat{m}_{0}^2 \, \sigma_0^4 \, + \, \lambda_0^2   ) \big)   $   &  $ \pi \, d_{01}  \, / \, b_{01} \, - \, 4 \, \pi^3 \, a_{01} \, \sigma_0^4  \, / \, \big( b_{01}  (4 \, \pi^2 \, a_{01}^2 \, \sigma_0^4 \,+ \,b_{01}^2 ) \big) $  \\
 \cline{1-3} 
 $\chi_0$  &   $  \pi^{-1 \, / \, 4} \, \sqrt{ \widehat{m}_{0}\, \sigma_0   \, / \,  \big( \cos(\alpha_0)\,  \widehat{m}_{0} \, \sigma_0^2 \,+ \, \imath \, \lambda_0 \big) }$  &   $\exp \left( - \imath \, \pi \, / \, 4 \right) \sqrt{ 2 \, \sqrt{\pi } \, \sigma_0   \, / \, \big( b_{01} \, - \,2  \, \imath \,\pi  \, a_{01}  \,\sigma_0^2 \big) }$ \\
 \hline
 \end{tabular}
 \renewcommand{\arraystretch}{2.1}
 \begin{tabular}{|m{2.0cm}|m{7.8cm}|m{6.2cm} |}
\cline{1-3} 
\multicolumn{3}{ |c|}{$ \Psi_{2, n}(x_2) \, = \, \chi_0 \, \chi_{1, n} \, \exp \left(\widetilde{A}_{1, n} \, x_2^2 \,+ \, \imath \, \widetilde{B}_{1, n} \, x_2^2 \,+ \,C_{1, n} \, x_2 \, + \,  \imath \,  D_{1, n} \, x_2 \, \right)$  }    \\ 
\cline{1-3} 
 $\widetilde{A}_{1, n}$ &  $  \widetilde{\beta}_{1, n}^2 \, \widehat{m}_{12}^2 \, \left(2 \,  A_0 \, \widetilde{\beta}_{1, n}^2 \, - \, 1\right)   \, / \, \big( 2 \, \widetilde{\zeta}_{1, n} \big) $  &   $ 2 \, \pi^2 \, \widetilde{\beta}_{1, n}^2 \, (2 \, A_{0} \,\widetilde{\beta}_{1, n}^2-1 )   \, / \, \widetilde{\zeta}_{1, n} $   \\
\cline{1-3} 
$\widetilde{B}_{1, n} $ &  \makecell[{{l}}]{$  \big( 2 \, B_0 \, \widetilde{\beta}_{1, n}^4 \, \widehat{m}_{12}^2 \, \cos (2 \, \alpha_1) \,+ \, \cos(\alpha_1) \, \lambda_1 \, \widehat{m}_{12}  \, \varrho_1 \big)   \, / \, \big(2 \, \widetilde{\zeta}_{1, n} \big)  $ \\
$  -  \, \widetilde{\beta}_{1, n}^4  \, \widehat{m}_{12}^3 \, \cos(\alpha_1) \, \sin ^2(\alpha_1)      \, / \, \big( 2 \, \lambda_1 \, \widetilde{\zeta}_{1, n} \big) $ } &  $\pi \,  \left(d_{12} \, \widetilde{\zeta}_{1, n} \,- \,4 \,\pi \, \widetilde{\beta}_{1, n}^4 \,(\pi \, a_{12} \,+ \, B_0 \, b_{12})\right) \,/ \, \big( b_{12} \, \widetilde{\zeta}_{1, n} \big)$  \\
\cline{1-3} 
    $ \chi_{1,n} $, $C_{1,n} $,  
     $D_{1,n} $    & \multicolumn{2}{ c|}{$\sqrt{\widetilde{\xi}_{1, n}} \,\mbox{exp} \big( \widetilde{p}_{1,1, n}\, x_{1, n}^2 \big)  $, $\,\,\,\,\,\,\,   \widetilde{\zeta}_{1, c, n} \, x_{1, n} $, $\,\,\,\,\,\,\, \widetilde{\zeta}_{1, d, n} \, x_{1, n} $}       \\
\hline
 \end{tabular}
 \renewcommand{\arraystretch}{2.1}
 \begin{tabular}{|m{2.0cm}|m{7.8cm}|m{6.2cm} |}
       \cline{1-3} 
   \multicolumn{3}{|c|}{ $ \Psi_{j+1,n}  (x_{j+1})  \, = \,  \chi_{0} \, \big( \prod_{k=1}^{j} \chi_{k,n} \big) \, e^{(\widetilde{A}_{j, n} \,+ \, \imath \, \widetilde{B}_{j, n})\, x_{j+1}^2 \,+ \, (C_{j,n}  \,+ \, \imath \, D_{j,n})\, x_{j+1}} $ for $ j \, \in \, [2, N -1]$ }            \\ 
       \cline{1-3} 
        \makecell[{{l}}]{$\widetilde{p}_{2, j, n}$,  $\widetilde{p}_{3, j, n}$, \\ $\widetilde{p}_{4, j, n}$, $\widetilde{p}_{5, j, n}$}    & \makecell[{{l}}]{$   \widetilde{\beta}_{j, n}^2  \,  \widetilde{p}_{3, j, n}   \, / \,   2    $, $ - \, \lambda_j  \, / \, \big( \imath \, \widetilde{\varsigma}_{j, n} \big) $,  \\ $\widetilde{\beta}_{j, n}^2 \widetilde{\zeta}_{j, c, n}$, $ -\widetilde{\beta}_{j, n}^2 \widetilde{\zeta}_{j, d, n} $ }     & \makecell[{{l}}]{$ \widetilde{\beta}_{j, n}^2 \,  \widetilde{p}_{3, j, n}     \, / \,   2  $, $ \widetilde{\varsigma}_{j, n}   \, / \,\widetilde{\zeta}_{j, n} $,  \\  $\widetilde{\beta}_{j, n}^2 \widetilde{\zeta}_{j, c, n}$,  $ - \,\widetilde{\beta}_{j, n}^2 \, \widetilde{\zeta}_{j, d, n} $ }\\ 
\cline{1-3} 
  $\widetilde{A}_{j, n}$   & $   \widetilde{\beta}_{j, n}^2 \, \widehat{m}_{j}^2 \, \left(2 \,\widetilde{A}_{j-1, n} \, \widetilde{\beta}_{j, n}^2 \, - \, 1\right)   \, / \, \big( 2 \, \widetilde{\zeta}_{j, n} \big)  $   &  $2 \,\pi ^2 \,\widetilde{\beta}_{j, n}^2 \, (2 \widetilde{A}_{j-1, n} \, \widetilde{\beta}_{j, n}^2\, - \, 1 ) \, / \, \widetilde{\zeta}_{j, n}$ \\ 
  \cline{1-3} 
  $\widetilde{B}_{j, n}$  &   \makecell[{{l}}]{$  \widehat{m}_{j} \big(  2 \, \widetilde{B}_{j-1, n} \, \widetilde{\beta}_{j, n}^4   \, \cos(2 \,\alpha_j)  \, \widehat{m}_{j}   \, +  \, \cos(\alpha_j)\lambda_j \widetilde{\varrho}_{j, n} \big)   \, / \, \big( 2\, \widetilde{\zeta}_{j, n}  \big)  $\\   $ -  \,\big( \widetilde{\beta}_{j, n}^4 \, \widehat{m}_{j}^3  \, \cos(\alpha_j) \, \sin^2(\alpha_j)   \big)   \, / \, \big(2 \, \lambda_j \, \widetilde{\zeta}_{j, n} \big)  $}   &   $  \frac{\pi   \left(d_{j, j+1} \,\widetilde{\zeta}_{j, n}\,-\,4 \,\pi \, \widetilde{\beta}_{j, n}^4 (\pi \, a_{j, j+1}\,+\,\widetilde{B}_{j-1, n} \, b_{j, j+1})\right)}{\big( b_{j, j+1}\, \widetilde{\zeta}_{j, n} \big)}          $ \\ 
\cline{1-3} 
  $C_{j,n} $, $D_{j,n} $  & \multicolumn{2}{ c|}{$ \widetilde{\zeta}_{j, c, n}  \, x_{j, n}  \, +  \, \widetilde{p}_{4, j, n} \, C_{j-1,n}    + \, \widetilde{p}_{5, j, n} \, D_{j-1,n} $, $ \, \, \, \, \,\,\,\,\, \widetilde{\zeta}_{j, d, n} \,  x_{j, n} \, -  \, \widetilde{p}_{5, j, n}\,  C_{j-1,n} \,   + \, \widetilde{p}_{4, j, n} \, D_{j-1,n}$  }  \\
\cline{1-3} 
$\chi_{j,n}$ &   \multicolumn{2}{ c|}{\makecell[{{c}}]{ $\sqrt{\widetilde{\xi}_{j, n}}   \, \mbox{exp}\big(\widetilde{p}_{1,j, n} \,  x_{j,n}^2 \big)  \, \times \, \mbox{exp}\big(\widetilde{p}_{2,j, n}   (C_{j-1,n} \, + \, \imath \, D_{j-1,n})^2 \big) \times  \,  \mbox{exp}\big(\widetilde{p}_{3, j, n} \,(C_{j-1,n} \, + \, \imath \, D_{j-1,n}) \,  x_{j,n}\big) $ }}\\
\hline
\end{tabular}
 \renewcommand{\arraystretch}{2.1}
 \begin{tabular}{|m{2.0cm}|m{7.8cm}|m{6.2cm} |}
          \cline{1-3} 
   \multicolumn{3}{|c|}{ The following variables defined for $ j \in [1, N -1]$ }            \\ 
         \cline{1-3} 
   $\lambda_j$ or $\widetilde{\lambda}_{j,n}$    &   $\hbar  \, t_{j, j+1}$       &    $ b_{j, j+1} \, (\widetilde{A}_{j-1, n} \,+ \, \imath \widetilde{B}_{j-1, n}) \, + \, \imath \,\pi \, a_{j, j+1}$    \\ 
\cline{1-3} 
 $\widetilde{p}_{1, j, n}$  &  \makecell[{{l}}]{$-  \big( 2 \, \lambda_j \, (\widetilde{A}_{j-1, n} \, + \, \imath \widetilde{B}_{j-1, n}) \, + \, \imath \, \cos(\alpha_j) \, \widehat{m}_{j} \big)   \, / \, \big( 2  \, \imath \, \widetilde{\varsigma}_{j, n} \big) $}   & $ \widetilde{\lambda}_{j,n} \,  (b_{j, j+1} \, - \,2 \, \widetilde{\beta}_{j, n}^2 \, \widetilde{\lambda}_{j,n}^{*}  )   \, / \, \widetilde{\zeta}_{j, n} $  \\ 
\cline{1-3} 
   $\widetilde{\varsigma}_{j, n}$,  $\widetilde{\xi}_{j, n}$ &   \makecell[{{l}}]{$    \widetilde{\beta}_{j, n}^2 \, \big( \cos(\alpha_j) \, \widehat{m}_{j}  \,+ \, 2  \,  \lambda_j  \,(\widetilde{B}_{j-1, n} \, - \, \imath \, \widetilde{A}_{j-1, n}) \big) \, + \, \imath \,  \lambda_j$, \\ $\,\widetilde{\beta}_{j, n}^2  \, \widehat{m}_{j} \, / \, \widetilde{\varsigma}_{j, n}$ }  & \makecell[{{l}}]{$b_{j, j+1}  \,  (b_{j, j+1} \, - \, 2  \, \widetilde{\beta}_{j, n}^2  \, \widetilde{\lambda}_{j,n}^{*}  )$,  \\ $ 2 \, \pi \, \widetilde{\beta}_{j, n}^2   \, / \, \big( \imath \, ( b_{j, j+1} \, - \, 2 \, \widetilde{\beta}_{j, n}^2  \, \widetilde{\lambda}_{j,n}) \big) $ }\\
\cline{1-3} 
  $\widetilde{\varrho}_{j, n}$ &  \multicolumn{2}{ c|}{$  4 \, \widetilde{\beta}_{j, n}^4  \,  (\widetilde{A}_{j-1, n}^2 \, + \, \widetilde{B}_{j-1, n}^2 ) \, - \, 4 \, \widetilde{A}_{j-1, n} \,  \widetilde{\beta}_{j, n}^2 \, + \,1$  } \\
\cline{1-3} 
  $\widetilde{\zeta}_{j, n}$ & \makecell[{{l}}]{ $  4  \, \widetilde{B}_{j-1, n} \, \widetilde{\beta}_{j, n}^4  \, \cos(\alpha_j) \, \lambda_j \, \widehat{m}_{j}    \, + \, \widetilde{\beta}_{j, n}^4 \, \cos^2(\alpha_j) \, \widehat{m}_{j}^2 \, + \, \lambda_j^2   \, \widetilde{\varrho}_{j, n}$ }&  \makecell[{{l}}]{ $b_{j, j+1}^2 \,  \widetilde{\varrho}_{j, n} $  \\ $ + \,4 \,\pi \, a_{j, j+1} \, \widetilde{\beta}_{j, n}^4  \, (\pi  \, a_{j, j+1} \, + \, 2  \, \widetilde{B}_{j-1, n} \,  b_{j, j+1})$  }\\
\cline{1-3} 
  $\widetilde{\zeta}_{j, c, n}$ & $  \widetilde{\beta}_{j, n}^2  \, \widehat{m}_{j}  \,  (2  \, \widetilde{B}_{j-1, n}  \, \lambda_j \, +  \, \cos(\alpha_j) \, \widehat{m}_{j} ) \, / \, \widetilde{\zeta}_{j, n}$ &  $ 4\, \pi \, \widetilde{\beta}_{j, n}^2 \,(\pi \,  a_{j, j+1} \, +  \, \widetilde{B}_{j-1, n} \, b_{j, j+1})  \, / \, \widetilde{\zeta}_{j, n} $  \\
  \cline{1-3} 
   $\widetilde{\zeta}_{j, d, n}$ &   $  \lambda_j \, \widehat{m}_{j} \, (2  \, \widetilde{A}_{j-1, n} \, \widetilde{\beta}_{j, n}^2 \, - \,1 ) \, / \, \widetilde{\zeta}_{j, n}$ & $2 \,\pi  \, b_{j, j+1} \,  (2 \, \widetilde{A}_{j-1, n} \, \widetilde{\beta}_{j, n}^2 \, - \, 1 ) \, / \, \widetilde{\zeta}_{j, n}$     \\
\hline
 \end{tabular}
\end{center}
\label{Table_Gaussian} 
\end{table*} 

\setlength\tabcolsep{2 pt}    
\renewcommand{\arraystretch}{2}
\begin{table*}[!t]
\caption{ Polynomial expressions in (\ref{Eq_example1}-\ref{Eq_example4}) for the case of $N = 3$ and Gaussian source with $\sigma_0$ ($b_{j, j+1} \neq 0$ for $j \in [0,2]$) }
\vspace{-0.2in}
\begin{center}
\footnotesize
%\begin{tabularx}{0.98\linewidth}{|c|m{3.2cm}|c|} 
\begin{tabular}{|m{1.4cm}|m{16cm}|} 
% Poly. &  Expression &  Poly. &  Expression & Poly. & Expression \\
 \hline 
 $pol_1 $  &   $\pi  \,  \Big(-2 \, \imath \, b_{01} \,  q_{7}  \, \pi  \,  \beta_{2}^2 \, - \, 2  \, \pi \,   \big(2  \, \pi \,   (a_{01} \,  q_{7} \, - \, b_{12} \,  q_{12})  \, \beta_{2}^2 \, + \, b_{12} \,  b_{23} \, \imath \, q_{19}\big)  \, \sigma_0^2 \, + \, b_{01} \,  b_{12}  \, b_{23}  \, q_{11}\Big)$  \\
  \hline 
   $pol_2 $  &    $2  \, \pi \,   \big(4 \,  \beta_{1}^2 \,  \pi ^2  \, (a_{01}  \, q_{7} \, - \, b_{12} \,  q_{12})  \, \beta_{2}^2 \, - \, a_{01}  \, b_{01}  \, b_{12}^2 \,  b_{23} \, + \, 2  \, b_{12} \, \imath \, \pi  \,  q_{17}\big)  \, \sigma _0^2 \, + \, b_{01}  \, \big(4  \, \beta_{1}^2 \, \imath \, \pi^2  \, q_{7}  \, \beta_{2}^2 \, + \, b_{01}  \, b_{12}^2  \, b_{23}  \, (- \imath  ) \, - \, 2  \, b_{12}  \, q_{23}  \, \pi \big)$       \\
 \hline 
 $pol_3$, $pol_4$  &  $- \, 2  \, b_{01}  \, b_{12}  \, b_{23}  \, \pi  \,  (2  \, a_{01} \,  \pi   \, \sigma_0^2 \, + \, b_{01} \, \imath )$, $\pi \,   \Big( b_{01} \,  \big(b_{01} \,  b_{12}  \, q_{12} \, - \, 2 \, \imath \, \beta_{1}^2  \, q_{7}  \, \pi  \, \big) \, - \, 2  \, \pi  \,  \big(2 \,  \pi   \, (a_{01}  \, q_{7} \, - \, b_{12}  \, q_{12})  \, \beta_{1}^2 \, + \, a_{01}  \, b_{01}  \, b_{12} \, \imath \, q_{12}\big)  \, \sigma _0^2\Big)$  \\
\hline 
 $pol_5 $  &    $ - \, 4  \, \beta_{2}^2 \,  b_{01}^2  \, b_{12}^3  \, b_{23} \,  \pi^2  \, q_{20}  \, \Big(b_{12}^2 \,  b_{01}^4 \, + \, 4  \, \beta_{1}^2  \, \pi ^2 \,  q_{13}  \, b_{01}^2 \, + \, 4  \, \pi^2  \, \sigma_0^2  \, \big(2  \, \beta_{1}^2  \, q_{14}  \, b_{01}^2 \, + \,  (4  \, \pi^2  \, q_{28}  \, \beta_{1}^2 \, + \, a_{01}^2  \, b_{01}^2  \, b_{12}^2)  \,  \sigma _0^2\big)\Big)$  \\
\hline 
 $pol_6 $  &   \makecell[{{l}}]{$\Big(4  \, b_{01}^2 \,  \pi^2  \, q_{11}^2  \, \beta_{1}^4 \, + \, b_{01}^4  \, b_{12}^2 \, + \, 4  \, \pi^2  \, \sigma_0^2  \, \big(2  \, \beta_{1}^2  \, b_{01}^2  \, b_{12}^2 \, + \, (4  \, \pi^2  \, q_{19}^2  \, \beta_{1}^4 \, + \, a_{01}^2  \, b_{01}^2 \,  b_{12}^2)  \, \sigma _0^2\big)\Big) $  \\$ \times  \,  \big( (16  \, \beta_{1}^4  \, \pi ^4  \, q_{7}^2  \, \beta_{2}^4 \, + \, b_{01}^2  \, b_{12}^4  \, b_{23}^2 \, + \, 4 b_{12}^2  \, \pi^2  \, q_{8} )  \, b_{01}^2 \, + \, 4  \, \pi^2  \, \sigma _0^2   \, (2  \, \beta_{1}^2 \,  b_{01}^2  \, q_{9}  \, b_{12}^2 \, + \, q_{22}  \, \sigma _0^2 )\big)$ } 
   \\
\hline  
 $pol_7 $, $pol_{10} $    &   $16  \, \beta_{1}^2  \, \beta_{2}^4  \, b_{01}  \, b_{12} \,  \pi^4  \,  (4    \, a_{01}  \, \pi^2  \, q_{19}  \, \sigma_0^4 \, + \, b_{01}^2 \,  q_{11} )  \, q_{29}$,  $ \, \imath \, \beta_{1}^2  \, \beta_{2}^2  \, b_{01}  \, b_{12}  \, \sigma _0   \, (b_{01} \, - \, 2 \, \imath \, a_{01}  \, \pi   \, \sigma _0^2 ) \big(b_{01}  \,  (b_{01}  \, b_{12} \, - \, 2 \, \imath \, \beta_{1}^2  \,  q_{11}  \,  \pi  ) \, - \, 2 \pi \,   q_{18}  \, \sigma_0^2\big)$ \\
\hline 
 $pol_8 $    &  $ - \, 8  \, \beta_{1}^2  \, \beta_{2}^2  \, b_{01}  \, b_{12}^2  \, b_{23} \,  \pi^3   \, (4  \, a_{01}  \, \pi^2  \, q_{19}  \, \sigma _0^4 \, + \, b_{01}^2  \, q_{11} ) \Big(b_{12}^2 \,  b_{01}^4 \, + \, 4  \, \beta_{1}^2  \, \pi ^2  \, q_{13}  \, b_{01}^2 \, + \, 4  \, \pi ^2  \, \sigma_0^2  \,  \big(2  \, \beta_{1}^2 \,  q_{14} \,  b_{01}^2 \, +  \, (4 \,  \pi^2  \, q_{28}  \, \beta_{1}^2 \, + \, a_{01}^2  \, b_{01}^2 \,  b_{12}^2 ) \,\sigma_0^2\big)\Big)$  \\
\hline 
 $pol_9 $,  $pol_{12} $      &   $- \, 8 \,  \beta_{2}^4  \, b_{01}^2  \, b_{12}^2  \, \pi^3  \, q_{20}  \, q_{29} $,  $2 \,  \beta_{2}^2  \, b_{12}^2  \, \pi^2  \, \big(- \, 4  \, \beta_{1}^2  \, q_{13}  \, \pi^2  \, b_{01}^2 \, - \, 8  \, \beta_{1}^2  \, \pi^2  \, q_{14}  \, \sigma_0^2  \, b_{01}^2 \, - \, 4 (4  \, \pi^4  \, q_{28} \,  \beta_{1}^2 \, + \, a_{01}^2  \, b_{01}^2  \, b_{12}^2  \, \pi^2 )  \, \sigma_0^4 \, - \, b_{01}^4  \, b_{12}^2\big)$   \\
\hline 
 $pol_{11} $  &   \makecell[{{l}}]{ $ 
 (2  \, a_{01} \,  \pi  \,  \sigma _0^2 \, + \, b_{01} \, \imath \, ) \big(2  \, \pi  \,  q_{15}  \, \sigma_0^2 \, + \, b_{01} \,  (2  \, \pi  \,  q_{11} \,  \beta_{1}^2 \, + \, b_{01}  \, b_{12} \, \imath   )\big)  $ \\$ \times\,  \Big(b_{01}  \, q_{16} \, - \, 2  \, \pi  \,  \big(4  \, \beta_{1}^2  \, \pi^2  \, (a_{01} \,  q_{7} \, - \, b_{12} \,  q_{12}) \,  \beta_{2}^2 \, - \, a_{01}  \, b_{01}  \, b_{12}^2  \, b_{23} \, + \, 2  \, b_{12} \, \imath \, \pi   \, q_{17}\big) \,  \sigma_0^2\Big)$  }\\
\hline 
 $pol_{13} $  &    $  (16  \, \beta_{1}^4  \, \pi^4  \, q_{7}^2  \, \beta_{2}^4 \, + \, b_{01}^2  \, b_{12}^4  \, b_{23}^2 \, + \, 4  \, b_{12}^2  \, \pi^2  \, q_{8} ) \,  b_{01}^2 \, + \, 4  \, \pi^2  \, \sigma_0^2   \, (2  \, \beta_{1}^2  \, b_{01}^2  \, q_{9}  \, b_{12}^2 \, + \, q_{22} \,  \sigma _0^2 )$  \\
 \hline 
 $pol_{14} $  &   \makecell[{{l}}]{$4  \, \pi^3  \, \sigma_0^2 \,  \Big(\sigma_0^2 \,  \big(a_{01}^2  \, b_{01}^2  \, b_{12}^4  \, b_{23}^2  \, d_{23} \, + \, 16  \, \pi^4  \, \beta_1^4  \, \beta_2^4  \, q_{4} \,  (a_{01}  \, q_{7} \, - \, b_{12}  \, q_{12}) \, + \, 4  \, \pi^2  \, b_{12}^2  \, q_{1}\big) \, + \, 2  \, \beta_1^2  \, b_{01}^2  \, b_{12}^2  \, \big(4  \, \pi^2  \, \beta_2^2  \, q_{6} \, + \, b_{12}^2 \,  b_{23}^2  \, d_{23}\big)\Big)$ \\ $+ \, \pi  \,  b_{01}^2   \, (16 \,  \pi^4  \, \beta_1^4  \, \beta_2^4  \, q_{5}  \, q_{7} \, + \, b_{01}^2 \,  b_{12}^4 \,  b_{23}^2 \,  d_{23} \, + \, 4  \, \pi^2  \, b_{12}^2  \, q_{2} )$ } \\
\hline
\hline
 \multicolumn{2}{|c|}{ \makecell[{{l}}]{
The functions $q_{j}$ for $j  \, \in  \, [1, 30]$ utilized while defining the polynomials  are defined as follows: \\
$q_{1}  \, \equiv  \, b_{12}^2 \, b_{23}^2  \, d_{23}  \, \beta_{1}^4 \, - \, 2  \, a_{01} \,  b_{12}  \, b_{23}^2  \, q_{11}  \, d_{23}  \, \beta_{1}^4 \, + \, a_{01}^2 \,  q_{2}$,  
$q_{2} \,  \equiv  \, a_{23}^2 \,  b_{01}^2 \,  b_{12}^2  \, d_{23} \,  \beta_{2}^4 \, - \, a_{23}  \, b_{01}^2  \, b_{12}  \, q_{27} \,  \beta_{2}^4 \, + \, b_{23} \,  q_{3}$, \\
$q_{3} \,  \equiv  \, b_{23}  \, d_{23} \,  q_{11}^2  \, \beta_{1}^4 \, + \, 2  \, \beta_{2}^2  \, b_{01}^2 \,  b_{23}  \, d_{23} \,  \beta_{1}^2 \, + \, \beta_{2}^4  \, b_{01}^2  \, d_{12}  \, q_{26}$, 
$q_{4} \,  \equiv  \, b_{12} \,  (b_{12} \, - \, q_{12} \,  d_{23}) \, - \, a_{01}  \, b_{12}  \, q_{11} \, + \, a_{01}  \, d_{23}  \, q_{7}$, \\
$q_{5}  \, \equiv \,  - \, b_{01}  \, b_{23}  \, d_{23} \, + \, a_{12}  \, b_{01}  \, q_{25} \, + \, b_{12} \,  d_{01}  \, q_{25}$, 
$q_{6}  \, \equiv  \,  (\beta_{1}^2 \,  b_{23}^2 \, + \, \beta_{2}^2  \, q_{12}^2 )  \, d_{23} \, - \, \beta_{2}^2  \, b_{12} \,  q_{12}$,   
$q_{7} \,  \equiv  \, - \, b_{01}  \, b_{23} \, + \, a_{12}  \, b_{01} \,  q_{12} \, + \, b_{12}  \, d_{01} \,  q_{12}$,  \\
$q_{8}  \, \equiv  \, a_{23}^2  \, b_{01}^2  \, b_{12}^2 \,  \beta_{2}^4 \, + \, 2  \, a_{23}  \, b_{01}^2 \,  b_{12}  \, b_{23} \,  d_{12} \,  \beta_{2}^4 \, + \, b_{23}^2 \,  q_{24}$,
$q_{9}  \, \equiv \,  4 \,  \pi ^2  \,  (\beta_{1}^2 \,  b_{23}^2 \, + \, \beta_{2}^2  \, q_{12}^2 )  \, \beta_{2}^2 \, + \, b_{12}^2 \,  b_{23}^2$,  
$q_{10} \,  \equiv  \, (b_{12}  \, q_{12} \, - \, a_{01}  \, q_{7})^2 $,\\
$q_{11}  \, \equiv \,  a_{12}  \, b_{01} \, + \, b_{12}  \, d_{01} $, 
$q_{12}  \, \equiv  \, a_{23} \,  b_{12} \, + \, b_{23}  \, d_{12} $,
$q_{13} \,  \equiv  \, b_{12}^2  \, d_{01}^2  \, \beta_{1}^2 \, + \, 2  \, a_{12}  \, b_{01}  \, b_{12} \,  d_{01}  \, \beta_{1}^2 \, +  \, (a_{12}^2  \, \beta_{1}^2 \, + \, \beta_{2}^2 ) \,  b_{01}^2 $, 
$q_{14}  \, \equiv 2  \, \beta_{1}^2 \,  \pi ^2  \, \beta_{2}^2 \, + \, b_{12}^2 $,\\
$q_{15}  \, \equiv  \, a_{01}  \, b_{01}  \, b_{12} \, - \, 2 \, \imath \, \beta_{1}^2  \, q_{19}  \, \pi$, 
$q_{16} \,  \equiv \,  - \, 4 \, \imath \, \beta_{1}^2  \, q_{7}  \, \pi ^2 \,  \beta_{2}^2 \, + \, b_{01} \,  b_{12}^2  \, b_{23} \, \imath \,+ \, 2  \, b_{12}  \, \pi   \, q_{23} $,
$q_{17}  \, \equiv  \, a_{01}  \, q_{23} \, - \, \beta_{1}^2  \, b_{12}  \, b_{23} $, 
$q_{18}  \, \equiv  \, 2  \, \pi  \,  q_{19} \,  \beta_{1}^2 \, + \, a_{01}  \, b_{01}  \, b_{12} \, \imath \, $,\\
$q_{19} \,  \equiv \,  a_{01}  \, a_{12} \,  b_{01} \, - \, b_{12} \, + \, a_{01}  \, b_{12}  \, d_{01} $, 
$q_{20}  \, \equiv \,  b_{01}^2 \, + \, 4  \, \pi^2 \,  \sigma_0^2  \,  (\beta_{1}^2 \, + \, a_{01}^2 \,  \sigma_0^2 ) $,
$q_{21}  \, \equiv \,  b_{12}^2  \, b_{23}^2  \, \beta_{1}^4 \, - \, 2  \, a_{01} \,  b_{12} \,  b_{23}^2 \,  q_{11}  \, \beta_{1}^4 \, + \, a_{01}^2  \, q_{8} $,\\
$q_{22}  \, \equiv  \, 16 \,  \beta_{1}^4  \, \pi^4  \, q_{10}  \, \beta_{2}^4 \, + \, a_{01}^2  \, b_{01}^2 \,  b_{12}^4  \, b_{23}^2 \, + \, 4  \, b_{12}^2 \,  \pi ^2  \, q_{21} $,  
$q_{23}  \, \equiv  \, b_{23}  \, q_{11}  \, \beta_{1}^2 \, + \, a_{23} \,  \beta_{2}^2  \, b_{01} \,  b_{12}+\beta_{2}^2 \,  b_{01} \,  b_{23} \,  d_{12}$, \\
$q_{24}  \, \equiv a_{12}^2 \,  b_{01}^2 \,  \beta_{1}^4 \, + \, b_{12}^2  \, d_{01}^2  \, \beta_{1}^4 \, + \, 2  \, a_{12} \,  b_{01}  \, b_{12} \,  d_{01} \,  \beta_{1}^4 \, + \, 2  \, \beta_{2}^2 \,  b_{01}^2 \,  \beta_{1}^2 \, + \, \beta_{2}^4 \,  b_{01}^2 \,  d_{12}^2$,  
$q_{25} \,  \equiv \,  b_{23}  \, d_{12} \,  d_{23} \, + \, b_{12}  \, (a_{23} \,  d_{23} \, - \, 1)  $, \\
$q_{26} \,  \equiv  \, b_{23}  \, d_{12}  \, d_{23} \, - \, b_{12} $,
$q_{27} \,  \equiv  \, b_{12} \, - \, 2  \, b_{23}  \, d_{12}  \, d_{23}  $,
$q_{28} \,  \equiv  \, q_{13} \,  a_{01}^2 \, - \, 2 \,  \beta_{1}^2  \, b_{12}  \, q_{11}  \, a_{01} \, + \, \beta_{1}^2 \,  b_{12}^2 $, \\
$q_{29}  \, \equiv  \, 4  \, b_{01}^2 \,  \pi^2  \, q_{11} \,  q_{7} \,  \beta_{1}^4 \, + \, b_{01}^4  \, b_{12}^2  \, q_{12} \, + \, 4  \, \pi^2 \,  \sigma_0^2  \,  (2  \, \beta_{1}^2  \, b_{01}^2  \, q_{12}  \, b_{12}^2 \, + \, q_{30}  \, \sigma_0^2 ) $ and finally
$q_{30}  \, \equiv \,  4 \, \pi^2  \,  q_{19} (a_{01}  \, q_{7} \, - \, b_{12} \,  q_{12})  \, \beta_{1}^4 \, + \, a_{01}^2 \,  b_{01}^2  \, b_{12}^2  \, q_{12} $.}}\\
\hline
 \end{tabular}
\end{center}
\label{Table_Example_Gaussian} 
\end{table*}

\subsection*{Matrix formulation for LCT system with Hermite-Gaussian sources}
\label{appc}

{\color{black} Iteration parameters utilized in (\ref{Eq_new_8}-\ref{Eq_new_12}) are presented in Table \ref{Table_HG}.} Some variables and constants used  in Table  {\color{black}\ref{Table_HG} (not defined previously or for Table \ref{Table_Gaussian})}  are the following:  $\widetilde{\tau}_{a, j, n} \,\equiv \, b_{j, j+1} \,\widetilde{u}_{j-1, j, n}  \,+ \,\imath \, \pi \,  a_{j, j+1} $ for $j \in [2, N-1]$, $\tau_{a, 1}\, \equiv  \,b_{12} \, {\color{black}u}_{01} \, + \,\imath \, \pi \,  a_{12} $ {\color{black}and} $\widetilde{\Gamma}_{1, n} \,= \,2 \, \widetilde{\beta}_{1, n}^2 \,b_{1, 2} \,{\color{black}g}_{0,1}^2 \, + \,\widetilde{\tau}_{1, n}$.  The formulation in (\ref{Eq_new_8}) is obtained by using the integral equality for Hermite polynomials (Section 16.5 in Ref. \citen{bateman1954tables}) in an iterative manner along the planes: 
\begin{align}
\begin{split}
\label{Eq_new_48}
\int dx \, \exp\bigg(\frac{-\,(x \,- \,y)^2}{2} \bigg) \, H_{l}\bigg(\frac{a \, x}{ \sqrt{2}} \bigg) = \sqrt{2 \, \pi} \, (1 - a^2)^{l \, / \, 2}    \, H_{l}\bigg(\frac{a \, y}{\sqrt{2 \, (1 - a^2)}} \bigg) &
\end{split}
\end{align}

The parameters   {\color{black}$\vec{\widetilde{\eta}}_{N-1, n}$ and $\widetilde{\gamma}_{j, j+1, n}$ in $ \vec{\widetilde{\gamma}}_{N-1, n}^T \equiv \big[ \widetilde{\gamma}_{12, n} \,\, \widetilde{\gamma}_{23, n} \,\, \hdots \,\,  \widetilde{\gamma}_{N-1, N, n} \big] $}  utilized in (\ref{Eq_new_9}) are defined  as follows {\color{black}by using the iterations in Table \ref{Table_HG}:} 
\begin{eqnarray}
\label{Eq_new_49}
\vec{\widetilde{\eta}}^T_{N-1, n}   \equiv    \begin{bmatrix}
\vec{\widetilde{\Xi}}_{N-2,n}^T  &   0
\end{bmatrix} \, \begin{bmatrix}
\widetilde{\mathbf{\Lambda}}_{N-2, n} &   \vec{0}_{N-2}\\
\vec{0}_{N-2}^T  & 0  \\
\end{bmatrix} +   \vec{\widetilde{\varepsilon}}_{N-1, n}^T  &&  
\end{eqnarray}
where the following are  defined: 
\begin{eqnarray}
\label{Eq_new_50_51_52}
 \widetilde{\mathbf{\Lambda}}_{N-2, n}    \equiv      \big[
\vec{G}_1    \,\,   \vec{G}_2  \,\,  \hdots \,\,   \vec{G}_{N-2}   
\big]^T; \hspace{0.7in} \vec{G}_k  \equiv    [ (\vec{\widetilde{\gamma}}_{k, n}^\star)^T  \, \,  \vec{0}_{N-2-k}^T   ]^T ; \hspace{0.7in}  \vec{\widetilde{\gamma}}_{k, n}^\star   \equiv    \big[ \prescript{k}{}{ \widetilde{\gamma}_{1, 2, n}^\star }    \,\, \prescript{k}{}{ \widetilde{\gamma}_{2, 3, n}^\star } \,\, \hdots \,\,  \prescript{k}{}{ \widetilde{\gamma}_{k, k+1, n}^\star } \big]^T  \hspace{0.0in} && \\
\label{Eq_new_53_54_55}
\vec{\widetilde{\varepsilon}}_{N-1, n}   \equiv    \big[ \widetilde{\varepsilon}_{12, n} \,\, \widetilde{\varepsilon}_{23, n} \,\, \hdots \,\,  \widetilde{\varepsilon}_{N-1, N, n}  \big]^T ; \hspace{0.1in} \vec{\widetilde{\Xi}}_{N-2,n}    \equiv   \big[ \widetilde{\Xi}_{12, n} \,\, \widetilde{\Xi}_{23, n} \,\, \hdots \,\,  \widetilde{\Xi}_{N-2, N-1, n} \big]^T ;  \hspace{0.1in}  \widetilde{\Xi}_{j, j+1, n}    \equiv   \widetilde{h}_{c, j+1, j+2, n}  \, \prod_{k = j+2}^{N-1} \widetilde{h}_{a, k, k+1, n} &&
\end{eqnarray}
where $\vec{\widetilde{\Xi}}_{N-2,n}^T$ {\color{black}is defined} for $N > 2$  and $ \widetilde{\gamma}_{j, j+1, n}  = \prescript{N-1}{}{ \widetilde{\gamma}_{j, j+1, n}^\star }$ while $\prescript{l}{}{ \widetilde{\gamma}_{j, j+1, n}^\star }$ and $\widetilde{\varepsilon}_{j, j+1, n} $ are defined as follows:
  \begin{eqnarray}
   \label{Eq_new_56_57}
 \prescript{l}{}{ \widetilde{\gamma}_{j, j+1, n}^\star }  \equiv \left \{
  \begin{aligned}
    &\widetilde{v}_{a, j, j+1, n} \, \prod_{k = j+1}^{l} \widetilde{\beta}_{k, n}^2 \widetilde{v}_{a, k, k+1, n}, &&   j \geq 2 \\
    & \widetilde{v}_{a, 12, n} \, \prod_{k = 2}^{l} \widetilde{\beta}_{k, n}^2 \widetilde{v}_{a, k, k+1, n}, && j = 1
  \end{aligned} \right. ;  \hspace{0.2in}  \,\,\,\,\,\,\,   \widetilde{\varepsilon}_{j, j+1, n} \equiv \left \{
  \begin{aligned}
    &\widetilde{h}_{b, j, j+1, n} \prod_{k = j+1}^{N-1} \widetilde{h}_{a, k, k+1, n}, &&   j \geq 2 \\
    &\widetilde{h}_{a, 12, n} \prod_{k = 2}^{N-1} \widetilde{h}_{a, k, k+1, n}, && j = 1
  \end{aligned} \right.\hspace{0.4in}   && 
  \end{eqnarray}
where $\widetilde{\gamma}_{12, n}  \equiv \widetilde{v}_{a, 12, n}$ and  $\widetilde{\varepsilon}_{12, n} \equiv \widetilde{h}_{a, 12, n}$ for $N =2$, and $\prescript{1}{}{ \widetilde{\gamma}_{1, 2, n}^\star} \equiv \widetilde{v}_{a, 12, n}$.  {\color{black}Finally,} $\widetilde{\mathbf{H}}_{N-1, n}^{LCT/HG}$ {\color{black}becomes equal} to the following {\color{black}by using the iterations in Table \ref{Table_HG}} :
\begin{eqnarray}
  \label{Eq_new_58}
 {\color{black}\widetilde{\mathbf{H}}_{N-1, n}^{LCT/HG} \, \equiv \,}  - \, \widetilde{\theta}_{12, n} \begin{bmatrix}
1 &   \vec{0}_{N-2}^T\\
\vec{0}_{N-2}  & \mathbf{0}_{N-2}  \\
\end{bmatrix}  
 + \, \sum_{j=2}^{N-1}  \begin{bmatrix}
\widetilde{\mathbf{H}}_{j, n}^{LCT/HG, \chi} &   \mathbf{0}_{j, N-1-j} \\
\mathbf{0}_{N-1-j, j}  & \mathbf{0}_{N-1-j }  \\ 
\end{bmatrix} \hspace{0.2in}  &&  
\end{eqnarray} 
where   $\widetilde{\mathbf{H}}_{k, n}^{LCT/HG, \chi}$ is the following:
\begin{align}
\begin{split} 
\label{Eq_new_59}
\widetilde{\mathbf{H}}_{k, n}^{LCT/HG, \chi}  \, \equiv \, & 
\widetilde{\theta}_{a, k, k+1, n} \begin{bmatrix}
\vec{\widetilde{\gamma}}_{k-1, n}^\star \, (\vec{\widetilde{\gamma}}_{k-1, n}^\star)^T &   \vec{0}_{k-1}\\
\vec{0}_{k-1}^T  & 0  \\
\end{bmatrix}  \,  + \, \widetilde{\theta}_{b, k, k+1, n} \begin{bmatrix}
\mathbf{0}_{k-1} &   \vec{\widetilde{\gamma}}_{k-1, n}^\star\\
\vec{0}_{k-1}^T  & 0  \\
\end{bmatrix}   \, + \, \widetilde{\theta}_{c, k, k+1, n} \begin{bmatrix}
\mathbf{0}_{k-1} &  \vec{0}_{k-1}\\
\vec{0}_{k-1}^T  & 1  \\
\end{bmatrix}
\end{split}
\end{align}

\setlength\tabcolsep{2 pt}    
\renewcommand{\arraystretch}{1.5}
\begin{table*}[!t]
\caption{Iteration parameters for FPI modeling of MPD with the kernel  $K_{LCT}^{(a,b,c,d)}$  for Hermite-Gaussian sources  ($b_{j, j+1} \neq 0$)}
\vspace{-0.2in}
\begin{center}
\footnotesize
%\begin{tabularx}{0.98\linewidth}{|c|m{3.2cm}|c|} 
  \begin{tabular}{|m{0.85cm}|m{2cm} | m{0.82cm}|m{5.65cm} |m{0.82cm}|m{5.30cm} |}
\cline{1-6} 
 \multicolumn{6}{|c|}{  $\Psi_0(x_0) \, = \, (2^{1/4} \, / \, \sqrt{W_0 \, 2^l \, l! }) \, \mbox{exp}\big(- \, \pi \, x_0^2 \, / \, W_0^2 \big) \, H_l( \sqrt{2 \, \pi} \, x_0 \, / \, W_0) $   and  $\Psi_1(x_1) \, = \,  \chi_{01} \exp \left(u_{01} \,x_1^2\right) H_{l}(g_{01} \, x_1)$ }    \\
 \cline{1-6} 
Symbol & Formula & Symbol & Formula   & Symbol & Formula \\ 
\cline{1-6} 
 $g_{01}$ &     $ \sqrt{\frac{2 \pi \, W_0^2}{a_{01}^2\, W_0^4\, + \, b_{01}^2}}$     &  $u_{01}$   & $-\frac{\pi \, W_0^2}{a_{01}^2 \, W_0^4 \,+ \, b_{01}^2} \, + \, \imath \,  \left(\frac{\pi  d_{01}}{b_{01}}-\frac{\left(\pi \,  W_0^2\right) \left(a_{01} \, W_0^2\right)}{b_{01} \left(a_{01}^2 \, W_0^4 \, + \, b_{01}^2\right)}\right) $  &  $\chi_{01}$  &   \vspace{0.05in} $ \frac{2^{1/4} \, \sqrt{W_0}}{\sqrt{2^l \, l!}} \,  \sqrt{\frac{a_{01} \, W_0^2 \, - \,\imath \,  b_{01}}{a_{01}^2 \,  W_0^4 \, + \, b_{01}^2}} \left(\frac{a_{01} W_0^2 \,- \,\imath \,  b_{01}}{\sqrt{a_{01}^2 \, W_0^4 \,+ \, b_{01}^2}}\right)^{l}$  \\
 \hline  
 \end{tabular} 
  \begin{tabular}{|m{0.85cm}|m{2.8cm} | m{0.7cm}|m{2.5cm} | m{0.8cm}|m{1.4cm} | m{0.8cm}|m{1.8cm} | m{0.7cm}|m{2.50cm} |}
\cline{1-10} 
  \multicolumn{10}{|c|}{$\Psi_{2,n} (x_2) \, = \, \chi_{01}  \, \chi_{12, n}  \, \exp \left(\widetilde{u}_{12, n} \, x_2^2 \, + \, v_{12, n} \, x_2\right) H_{l}(\widetilde{g}_{12, n} x_2 \, + \,  h_{12, n})$} \\
\cline{1-10} 
 & Formula &  & Formula &  & Formula &  & Formula &  & Formula  \\ 
 \cline{1-10} 
$\widetilde{\tau}_{1, n}$ &  $ - \, b_{12} \, + \,2 \, \widetilde{\beta}_{1, n}^2  \,\tau_{a, 1}$ & $\widetilde{u}_{12, n}$ & \vspace{0.05in} $  \frac{2 \, \pi ^2\,  \widetilde{\beta}_{1, n}^2 \,+ \, \imath \,  \pi \,  d_{12} \, \widetilde{\tau}_{1, n}}{b_{12}\, \widetilde{\tau}_{1, n}}$ & $\widetilde{v}_{a, 12, n} $ & $  \frac{\imath \, 2  \, \pi }{\widetilde{\tau}_{1, n}}$ &  $\widetilde{h}_{a, 12, n}$ & $  - \, \frac{b_{12} \, g_{01}}{\sqrt{\widetilde{\tau}_{1, n}} \, \sqrt{ \widetilde{\Gamma}_{1, n}}} $ & $\widetilde{\theta}_{12, n}$ & $ \frac{b_{12} \, u_{01} \, + \, \imath \, \pi  \, a_{12}}{\widetilde{\tau}_{1, n}}$  \\
 \cline{1-10} 
$ \widetilde{\chi}_{a, 12, n}$ & \vspace{0.05in} $ \sqrt{2 \,\pi } \, \sqrt{\frac{\imath \, \widetilde{\beta}_{1, n}^2}{\widetilde{\tau}_{1, n}}} \left(\frac{\widetilde{\Gamma}_{1, n} }{\widetilde{\tau}_{1, n}}\right)^{l/2}$ &  $\widetilde{g}_{12, n} $ &  $\frac{2 \, \imath \,  \pi \, \widetilde{\beta}_{1, n}^2 \, g_{01}}{\sqrt{\widetilde{\tau}_{1, n}} \sqrt{ \widetilde{\Gamma}_{1, n} }}$ & $v_{12, n}$ & $ \widetilde{v}_{a, 12, n} \, x_{1, n} $ & $h_{12, n}$ & $\widetilde{h}_{a, 12, n} \, x_{1, n}$ & $\chi_{12, n}$ & $\widetilde{\chi}_{a, 12, n} \, e^{-\, \widetilde{\theta}_{12, n} \,  x_{1, n}^2 }$ \\
   \hline
 \end{tabular}
   % \vspace{0.1in}
 \renewcommand{\arraystretch}{1.5}
 \begin{tabular}{|m{1.15cm}|m{2.35cm} | m{1.2cm}|m{2.8cm} | m{1.15cm}|m{3.0cm} | m{1.12cm}|m{2.4cm}   |}
\cline{1-8} 
   \multicolumn{8}{|c|}{ $ \Psi_{j+1,n}  (x_{j+1})  \, = \, \chi_{01} \, \big( \prod_{i = 1}^{j} \chi_{i,i+1, n}  \big)\exp \big(\widetilde{u}_{j, j+1, n} \, x_{j+1}^2 \, + \,v_{j, j+1, n} \, x_{j+1} \big) \,H_{l}(\widetilde{g}_{j, j+1, n} \, x_{j+1} \, + \, h_{j, j+1,  n})      $ for $ j \in [2, N -1]$}  \\
\cline{1-8} 
 & Formula &  & Formula &  & Formula &  & Formula     \\
 \cline{1-8} 
 $\widetilde{\tau}_{j, n}$ & \vspace{0.05in} \makecell[{{l}}]{$- \, b_{j, j+1} $ \\ $ + \,2 \, \widetilde{\beta}_{j, n}^2  \, \widetilde{\tau}_{a, j, n}  $} & $\widetilde{u}_{j, j+1, n}$  &  $ \frac{2 \,\pi^2  \, \widetilde{\beta}_{j, n}^2 \, + \, \imath \,  \pi  \,  d_{j, j+1} \widetilde{\tau}_{j, n}}{b_{j, j+1} \widetilde{\tau}_{j, n}}$ & $\widetilde{v}_{a, j, j+1, n}$ & $  \frac{\imath \,  2 \, \pi }{\widetilde{\tau}_{j, n}}$ & $\widetilde{\Gamma}_{j, n}$  &   \makecell[{{l}}]{$2 \, \widetilde{\beta}_{j, n}^2 b_{j, j+1} \widetilde{g}_{j-1,j, n}^2 $ \\ $+ \,\widetilde{\tau}_{j, n}$} \\  
 \cline{1-8} 
  $\widetilde{\theta}_{a, j, j+1, n}$  &  \vspace{0.05in} $  -\frac{\widetilde{\beta}_{j, n}^2 \, b_{j, j+1}}{2 \, \widetilde{\tau}_{j, n}} $ & $\widetilde{\theta}_{b, j, j+1, n}$ & $ -\, \frac{b_{j, j+1}}{\widetilde{\tau}_{j, n}}$ & $\widetilde{\theta}_{c, j, j+1, n}$ & $  - \, \frac{b_{j, j+1} \,  \widetilde{u}_{j-1, j, n} \, + \, \imath \,  \pi  a_{j, j+1}}{\widetilde{\tau}_{j, n}} $ &   $ \widetilde{h}_{c, j, j+1, n}$ & \vspace{0.05in} $ -\, \frac{\widetilde{\beta}_{j, n}^2 \, b_{j, j+1} \, \widetilde{g}_{j-1, j, n}}{\sqrt{\widetilde{\tau}_{j, n}} \sqrt{\widetilde{\Gamma}_{j, n}}}$     \\
  \cline{1-8} 
$\widetilde{h}_{a, j, j+1, n} $ & $ \frac{\sqrt{\widetilde{\tau}_{j, n}}}{\sqrt{\widetilde{\Gamma}_{j, n}}}   $ & $\widetilde{h}_{b, j, j+1, n}$ & $ - \, \frac{b_{j, j+1} \widetilde{g}_{j-1, j, n}}{\sqrt{\widetilde{\tau}_{j, n}} \sqrt{\widetilde{\Gamma}_{j, n}}} $ &  $ \widetilde{\chi}_{a, j, j+1, n}$ & \vspace{0.05in} $  \sqrt{2 \, \pi } \sqrt{- \, \frac{\widetilde{\beta}_{j, n}^2}{\widetilde{\tau}_{j, n}}} \left(\frac{\widetilde{\Gamma}_{j, n}}{\widetilde{\tau}_{j, n}}\right)^{l/2} $ &  $\widetilde{g}_{j, j+1, n}$  & $  \frac{2 \, \imath \,  \pi  \, \widetilde{\beta}_{j, n}^2\,  \widetilde{g}_{j-1, j, n}}{\sqrt{\widetilde{\tau}_{j, n}} \sqrt{\widetilde{\Gamma}_{j, n}}}  $      \\
  \cline{1-8} 
  $h_{j, j+1,  n} $   &   \makecell[{{l}}]{$h_{j-1, j, n}  \, \widetilde{h}_{a, j, j+1, n} $ \\ $+ \, \widetilde{h}_{b, j, j+1, n} \, x_{j, n}  \, $ \\ $ + \, v_{j-1, j, n} \, \widetilde{h}_{c, j, j+1, n}   $} &  $v_{j, j+1, n} $  & \makecell[{{l}}]{ $ \widetilde{v}_{a, j, j+1, n} $ \\ $  \times \, (x_{j, n}  +\, \widetilde{\beta}_{j, n}^2 \, v_{j-1, j, n})   $ }& $\chi_{j, j+1, n} $  &  \multicolumn{3}{ l|}{  \makecell[{{l}}]{$  \widetilde{\chi}_{a, j, j+1, n} \exp \bigg( v_{j-1, j, n} \,\widetilde{\theta}_{b, j, j+1, n} \,x_{j, n}  $ \\  \hspace{0.7in} $+\,\widetilde{\theta}_{c, j, j+1, n} \,x_{j, n}^2  $ \\  \hspace{0.7in} $  + \, v_{j-1, j, n}^2 \widetilde{\theta}_{a, j, j+1, n} \, - \, \frac{\imath \,  \pi }{4} \bigg)    $ }  }    \\
  \hline 
 \end{tabular}
\end{center}
\label{Table_HG} 
\end{table*}

\section*{Competing interests}
The author  declares no competing interests.

%\bibliography{PhotonicQPC}

\end{document}